\documentclass[english,aps,manuscript,11pt,aps,letter,superscriptaddress,nofootinbib,pre]{revtex4}
\usepackage[T1]{fontenc}
\usepackage[latin9]{inputenc}
\usepackage{babel}
\usepackage{amsmath}
\usepackage{amssymb}
\usepackage{graphicx}
\usepackage[unicode=true,pdfusetitle,
 bookmarks=true,bookmarksnumbered=false,bookmarksopen=false,
 breaklinks=false,pdfborder={0 0 1},backref=false,colorlinks=false]
 {hyperref}

\makeatletter
\@ifundefined{textcolor}{}
{%
 \definecolor{BLACK}{gray}{0}
 \definecolor{WHITE}{gray}{1}
 \definecolor{RED}{rgb}{1,0,0}
 \definecolor{GREEN}{rgb}{0,1,0}
 \definecolor{BLUE}{rgb}{0,0,1}
 \definecolor{CYAN}{cmyk}{1,0,0,0}
 \definecolor{MAGENTA}{cmyk}{0,1,0,0}
 \definecolor{YELLOW}{cmyk}{0,0,1,0}
}

\usepackage[T1]{fontenc}
\usepackage{ae,aecompl}
\usepackage{lmodern}
\usepackage{amsfonts}
\usepackage{bm}
\usepackage{xfrac}
\usepackage[normalem]{ulem}
\usepackage{siunitx}
\usepackage{etoolbox}

\makeatother

\begin{document}
\global\long\def\V#1{\boldsymbol{#1}}%
\global\long\def\M#1{\boldsymbol{#1}}%
\global\long\def\Set#1{\mathbb{#1}}%

\global\long\def\D#1{\Delta#1}%
\global\long\def\d#1{\delta#1}%

\global\long\def\norm#1{\left\Vert #1\right\Vert }%
\global\long\def\abs#1{\left|#1\right|}%

\global\long\def\grad{\M{\nabla}}%
\global\long\def\avv#1{\langle#1\rangle}%
\global\long\def\av#1{\left\langle #1\right\rangle }%

\global\long\def\myhalf{\sfrac{1}{2}}%
\global\long\def\mythreehalves{\sfrac{3}{2}}%

\global\long\def\lapl{{\nabla^{2}}}%
 
\global\long\def\dlapl{{\nabla_{\mathrm{d}}^{2}}}%
 
\global\long\def\divg{{\V{\nabla}\cdot}}%
\global\long\def\ddivg{{\V{\nabla}_{\mathrm{d}}\cdot}}%

\global\long\def\ddt{{\frac{d}{dt}}}%
 
\global\long\def\pddt{{\frac{\partial}{\partial t}}}%
\global\long\def\paren#1{{(#1)}}%

\global\long\def\Var#1{{\mathrm{Var}\left[#1\right]}}%
 
\global\long\def\Cov#1#2{{\mathrm{Cov}\left[#1,#2\right]}}%
 
\global\long\def\equald{{\stackrel{\mathrm{d}}{=}}}%

\global\long\def\chisub{\chi^{\mathrm{sub}}}%
 
\global\long\def\chisubb{\M{\chi}^{\mathrm{sub}}}%

\title{Fluctuating hydrodynamics of electrolytes at electroneutral scales}
\author{Aleksandar Donev}
\email{donev@courant.nyu.edu}

\affiliation{Courant Institute of Mathematical Sciences, New York University, New
York, NY, 10012}
\author{Andrew J. Nonaka}
\affiliation{Center for Computational Science and Engineering, Lawrence Berkeley
National Laboratory, Berkeley, CA, 94720}
\author{Changho Kim}
\affiliation{Center for Computational Science and Engineering, Lawrence Berkeley
National Laboratory, Berkeley, CA, 94720}
\affiliation{Department of Applied Mathematics, University of California, Merced,
CA, 95343}
\author{Alejandro L. Garcia}
\affiliation{Department of Physics, San Jose State University, San Jose, CA, 95192}
\author{John B. Bell}
\affiliation{Center for Computational Science and Engineering, Lawrence Berkeley
National Laboratory, Berkeley, CA, 94720}
\begin{abstract}
At mesoscopic scales electrolyte solutions are modeled by the fluctuating
generalized Poisson-Nernst-Planck (PNP) equations {[}J.-P. Péraud
\emph{et al.}, Phys. Rev. F, 1(7):074103, 2016{]}. However, at length
and time scales larger than the Debye scales, electrolytes are effectively
electroneutral, and the charged-fluid PNP equations become too stiff
to solve numerically. Here we formulate the isothermal incompressible
equations of fluctuating hydrodynamics for reactive multispecies mixtures
involving charged species in the electroneutral limit, and design
a numerical algorithm to solve these equations. Our model does not
assume a dilute electrolyte solution but rather treats all species
on an equal footing, accounting for cross-diffusion and non-ideality
using Maxwell-Stefan theory. By enforcing local electroneutrality
as a constraint, we obtain an elliptic equation for the electric potential
that replaces the Poisson equation in the fluctuating PNP equations.
We develop a second-order midpoint predictor-corrector algorithm to
solve either the charged-fluid or electroneutral equations with only
a change of the elliptic solver. We use the electroneutral algorithm
to study a gravitational fingering instability, triggered by thermal
fluctuations, at an interface where an acid and base react to neutralize
each other. Our results demonstrate that, because the four ions diffuse
with very different coefficients, one must treat each ion as an individual
species, and cannot treat the acid, base, and salt as neutral species.
This emphasizes the differences between electrodiffusion and classical
Fickian diffusion, even at electroneutral scales.
\end{abstract}
\maketitle

\section{Introduction}

Better understanding of transport phenomena in electrolytes is important
for studying both naturally occurring and man-made systems at small
scales. Living cells rely strongly on membrane potentials and the
electrodiffusion of ions. Batteries and fuel cells also rely on ionic
transport. In both of these examples the length and time scales involved
are intractable for molecular dynamics. A more efficient and tractable
numerical approach for mesoscopic fluids is fluctuating hydrodynamics
(FHD), which extends conventional hydrodynamics by including a random
component to the dissipative fluxes in a manner consistent with irreversible
thermodynamics and the fluctuation-dissipation theorem. Access to
tools to model systems involving complex electrolyte mixtures with
the inclusion of their inherent statistical fluctuations would not
only increase our understanding of cellular mechanisms, but also provide
a path towards better design tools for bio-engineering applications.

In our prior work \cite{LowMachElectrolytes} we formulated a \emph{charged-fluid}
form of the equations of fluctuating hydrodynamics and developed associated
algorithms for electrolyte mixtures containing an arbitrary number
of ionic or neutral species. Our formulation combined a generalized
fluctuating Poisson-Nernst-Planck (PNP) equation based on the Maxwell-Stefan
formulation of electrodiffusion with the fluctuating low Mach number
Navier-Stokes (NS) equation for the fluid flow. In that formulation,
the fluid is considered to be a mixture of incompressible but miscible
components (species), each with its own density, and it is not necessary
to distinguish a single species as a solvent \footnote{This formulation is also useful for modeling ionic liquids.}.
For very dilute electrolyte solutions, in the absence of fluctuations
the deterministic formulation reverts to the classical PNP equations
for the composition, coupled to an incompressible NS equation for
the fluid velocity. In recent work \cite{RenormalizationElectrolytes,RenormalizationElectrolytesTernary}
we have demonstrated that the addition of thermal fluctuations renormalizes
the PNP equations to reproduce the Debye-Hückel-Onsager theory for
dilute solutions.

The charged-fluid formulation is designed for simulations where the
spatial grid resolves the Debye length $\lambda_{D}$, which is typically
on the order of a few to tens of nanometers. In particular, the time
step size in the algorithm used in \cite{LowMachElectrolytes} was
limited by $\tau_{D}=\lambda_{D}^{2}/D$ (see Eq. (86) in \cite{LowMachElectrolytes}),
where $D$ is a typical diffusion coefficient. In many practical applications
one is interested in modeling bulk electrolytes at length scales much
larger than the Debye length, over diffusive time scales much longer
than $\tau_{D}$. At such scales, the electrolyte is effectively \emph{electroneutral},
and electrodiffusion is described by the electroneutral limit of the
PNP equations \cite{Electrophysiology_Peskin_Review,ElectroneutralAsymptotics_Yariv}.
In this paper, we formulate the electroneutral limit of the generalized
fluctuating PNP equations and develop a numerical method to solve
these equations. In the electroneutral limit, the evolution is constrained
to preserve charge neutrality by replacing the standard Poisson equation
for the electric field with a variable coefficient elliptic equation.
Thus, with only a change of an elliptic equation solver, our algorithm
can switch from charged-fluid to electroneutral, allowing us to use
the same code to study a broad range of length and time scales. Implicit
in a coarse-grained description like FHD is the assumption that each
cell (coarse-graining volume) contains sufficiently many ions to justify
neglecting the discrete particle nature of molecules. While this assumption
is problematic for charged-fluid FHD except for dilute solutions (for
which the Debye length is large compared to the inter-ion spacing),
in electroneutral FHD the cell dimensions are much larger than the
Debye length and therefore typically contain a large number of ions
even for dense solutions (for which the Debye length is comparable
or smaller than the inter-ion spacing).

Additionally, in this work we incorporate chemical reactions in the
charged-fluid and electro-neutral formulations/algorithms following
our prior work on non-ionic mixtures \cite{FluctReactFHD}. In the
approach developed in \cite{FluctReactFHD}, fluctuating chemistry
is treated using a discrete Chemical Master Equation (CME) formulation,
while hydrodynamic transport including mass and momentum diffusion
is treated using a fluctuating hydrodynamics semi-continuum formulation.
Our numerical algorithm is a modification of the algorithm developed
in \cite{FluctReactFHD} to replace diffusion by electrodiffusion
for both formulations.

In \cite{FluctReactFHD}, we modeled recent experiments \cite{DiffusiveInstability_AcidBase}
studying a gravity-driven instability of a front where an acid (HCl)
and a base (NaOH) neutralize each other to form a salt (NaCl). In
these prior simulations, we followed the literature \cite{BuoyancyInstabilities_Chemistry,InstabilityRT_Chemistry,DiffusiveInstability_AcidBase}
and modeled the acid, base, and salt as neutral species (HCl, NaOH,
and NaCl); we will refer to this as the \emph{ambipolar approximation}.
In reality, however, these species are all strong electrolytes and
disassociate into ions ($\mathrm{H}^{+}$, $\mathrm{OH}^{-}$, $\mathrm{Na}^{+}$,
and $\mathrm{Cl}^{-}$). It is well-known that electrodiffusion can
be very different than ordinary diffusion because of the strong coupling
of the motions of the ions via the electric fields they generate;
for example, an ionic species can diffuse against its own concentration
gradient \cite{ElectrolytesMS_Review}. In this work, we use the electroneutral
formulation to model the fingering instability at an HCl/NaOH front
but treating each ion as a separate charged species. This avoids uncontrolled
approximations and allows us to assess the quantitative accuracy of
the ambipolar approximation in a multispecies electrolyte.

We begin by formulating the stochastic partial differential equations
of fluctuating hydrodynamics for electrolytes in Section \ref{sec:Formulation}.
We first review the charged-fluid formulation in which the Debye length
is resolved in Section \ref{subsec:Electrocharged}, and then formulate
the electroneutral equations in Section \ref{subsec:Electroneutral}.
We also discuss the spectra of concentration fluctuations at thermodynamic
equilibrium for both charged-fluid and electroneutral formulations
in Section \ref{subsec:Fluctuations}. We present a second-order predictor-corrector
algorithm for both formulations in Section \ref{sec:NumericalAlgorithm}.
The methodology is applied to study a fingering instability at an
acid-base front in Section \ref{sec:AcidBase}. We conclude with some
directions for future research in Section \ref{sec:Conclusions}.

\section{\label{sec:Formulation}Charged-fluid and Electroneutral Fluctuating
Electrohydrodynamics}

We consider an isothermal isobaric mixture of $N_{s}$ species and
use the following notation. Vectors (both in the geometrical and in
the linear algebra sense), matrices (and tensors), and operators are
denoted with bold letters. The mass density of species $s$ is denoted
with $\rho_{s}$ and its number density with $n_{s}$, giving the
total mass density $\rho=\sum_{s=1}^{N_{s}}\rho_{s}$ and total number
density $n=\sum_{s=1}^{N_{s}}n_{s}$. The mass fractions are denoted
with $\V w$, where $w_{s}=\rho_{s}/\rho$, while the number or mole
fractions are denoted with $\V x$, where $x_{s}=n_{s}/n$; both the
mass and number fractions sum to unity. One can transform between
mass and number fractions by $x_{s}=\bar{m}w_{s}/m_{s}$, where $m_{s}$
is the molecular mass of species $s$ and the mixture-averaged molecular
mass is 
\[
\bar{m}=\frac{\rho}{n}=\left(\sum_{s=1}^{N_{s}}\frac{w_{s}}{m_{s}}\right)^{-1}.
\]
A diagonal matrix whose diagonal is given by a vector is denoted by
the corresponding capital letter; for example, $\M W$ is a diagonal
matrix with entries $\V w$ and $\V M$ is a diagonal matrix of the
molecular masses $\V m$.

The charges per unit mass are denoted by $\V z$ with $z_{s}=V_{s}e/m_{s}$,
where $e$ is the elementary charge and $V_{s}$ is the valence of
species $s$. The total density of free charges is thus
\[
Z=\sum_{s=1}^{N_{s}}\rho_{s}z_{s}=\rho\V z^{T}\V w.
\]
For an ideal solution the Debye length is given by
\begin{equation}
\lambda_{D}=\left(\frac{\epsilon k_{B}T}{\rho\V z^{T}\M W\M M\V z}\right)^{1/2}=\left(\frac{\epsilon k_{B}T}{\sum_{s=1}^{N_{s}}\rho w_{s}m_{s}z_{s}^{2}}\right)^{1/2},\label{DebyeLengthEqn}
\end{equation}
where $\epsilon$ is the dielectric permittivity of the mixture, $k_{B}$
is Boltzmann's constant, and $T$ is the temperature.

\subsection{\label{subsec:Electrocharged}Charged-fluid Formulation}

In this section we review the fluctuating hydrodynamics equations
for an electrolyte mixture, following the notation in \cite{LowMachElectrolytes}.
Unlike this prior work, and following \cite{FluctReactFHD}, here
we make a Boussinesq approximation and assume that the density of
the mixture changes only weakly with composition, $\rho\approx\rho_{0}$.
This allows us to use an incompressible approximation of the momentum
equation, which greatly simplifies the construction of a numerical
algorithm \cite{FluctReactFHD}. The dependence of the density on
composition is only taken into account in the gravity forcing term.
This Boussinesq approximation is certainly valid for moderately dilute
electrolyte solutions. We neglect the effects of thermodiffusion and
barodiffusion on mass transport and assume constant temperature $T$
and thermodynamic pressure $P$.

The incompressible equations of fluctuating hydrodynamics for an isothermal
reactive electrolyte mixture can be obtained by combining terms given
in \cite{LowMachElectrolytes} with those given in \cite{FluctReactFHD}.
Here we summarize the resulting equations.

\subsubsection{Quasi-electrostatic Poisson Equation}

In the electroquasistatic approximation (magnetic effects are neglected),
the electric potential $\Phi(\V r,t)$ satisfies the Poisson equation
\begin{equation}
\grad\cdot\left(\epsilon\V E\right)=-\grad\cdot\left(\epsilon\grad\Phi\right)=Z,\label{eq:PoissonEq}
\end{equation}
where the electric field is $\V E=-\grad\Phi$, and the dielectric
permittivity $\epsilon\left(\V w\right)$ can, in principle, depend
on composition. The boundary conditions for $\Phi$ are standard Neumann
conditions for dielectric boundaries or Dirichlet conditions for metallic
boundaries.

The presence of charges and electric fields leads to a nonzero Lorentz
force in the momentum equation given by the divergence of the Maxwell
stress tensor $\M{\sigma}_{M}=\epsilon\left(\V E\V E^{T}-\V E^{T}\V E\M I/2\right)$,
\[
\V f_{E}=\grad\cdot\M{\sigma}_{M}=Z\V E-\frac{E^{2}}{2}\grad\epsilon.
\]
In this work we assume that the permittivity is constant, which reduces
the Lorentz force to $\V f_{E}=Z\V E$. Using the Poisson equation
(\ref{eq:PoissonEq}), we can rewrite this in the equivalent form
\[
\V f_{E}=\grad\cdot\left(\epsilon\V E\right)\V E=[\grad\cdot\left(\epsilon\grad\Phi\right)]\grad\Phi,
\]
which is suitable for both the charged-fluid and the electroneutral
formulations \cite{ElectroneutralAsymptotics_Yariv}. By contrast,
as we explain later, the traditional form $\V f_{E}=Z\V E$ cannot
be used in the electroneutral limit since formally $Z\rightarrow0$
but the Lorentz force does \emph{not} go to zero in this limit \cite{ElectroneutralAsymptotics_Yariv}.

\subsubsection{Momentum Equation}

In the Boussinesq approximation, $\rho=\rho_{0}$ and conservation
of momentum gives the fluctuating incompressible Navier-Stokes equations
\begin{align}
\frac{\partial\left(\rho\V v\right)}{\partial t}+\grad\pi & =-\divg(\rho\V v\V v^{T})+\divg(\eta\bar{\grad}\V v+\M{\Sigma})+\grad\cdot\left(\epsilon\grad\Phi\right)\grad\Phi+\V f,\label{ddtvel}\\
\divg\V v & =0.\label{divvel}
\end{align}
Here, $\V v(\V r,t)$ is the fluid velocity, $\pi(\V r,t)$ is the
mechanical pressure (a Lagrange multiplier that ensures the velocity
remains divergence free), $\eta(\V w)$ is the viscosity, $\bar{\grad}=\grad+\grad^{T}$
is a symmetric gradient, and $\M{\Sigma}$ is the stochastic momentum
flux. The buoyancy force $\V f(\V w,t)$ is a problem-specific function
of $\V w(\V r,t)$ and can also be an explicit function of time.

Based on the fluctuation-dissipation relation, the stochastic momentum
flux $\M{\Sigma}$ is modeled as 
\begin{equation}
\M{\Sigma}=\sqrt{\eta k_{B}T}\left[\M{\mathcal{Z}}^{\text{mom}}+\left(\M{\mathcal{Z}}^{\text{mom}}\right)^{T}\right],
\end{equation}
where $\M{\mathcal{Z}}^{\text{mom}}(\V r,t)$ is a standard Gaussian
white noise tensor field with uncorrelated components having $\delta$-function
correlations in space and time.

The two physical boundary conditions for the charged-fluid equations
that we consider here are the \emph{no-slip} condition $\V v=\V 0$
on the boundary, and the \emph{free-slip} boundary condition, 
\begin{equation}
v_{n}=\V v\cdot\V n=0\quad\text{and}\quad\frac{\partial v_{n}}{\partial\V{\tau}}+\frac{\partial\V v_{\tau}}{\partial n}=\V 0,\label{eq:free_slip_BCs}
\end{equation}
where $\V n$ is the unit vector normal to the boundary, $\V{\tau}$
is a unit vector tangential to the boundary, $\V{\tau}\cdot\V n=0$,
and $\V v_{\tau}$ is the tangential component of the velocity.

\subsubsection{Species Equations}

Conservation of mass for each species gives the dynamics of the composition
of the mixture,
\begin{equation}
\frac{\partial\left(\rho w_{s}\right)}{\partial t}=-\divg(\rho w_{s}\V v)-\divg\V F_{s}+m_{s}\Omega_{s},\label{eq:ddtw}
\end{equation}
where we remind the reader that in the Boussinesq approximation density
is constant, i.e., $\rho=\rho_{0}$. The total diffusive mass flux
$\V F_{s}$ of species $s$ is composed of a dissipative flux $\overline{\V F}_{s}$
and fluctuating flux $\widetilde{\V F}_{s}$, 
\begin{equation}
\V F_{s}=\overline{\V F}_{s}+\widetilde{\V F}_{s},
\end{equation}
and $\Omega_{s}$ is a source term representing stochastic chemistry.
Note that by summing up (\ref{eq:ddtw}) over all species we recover
(\ref{divvel}) since $\sum_{s}\V F_{s}=\V 0$ and $\sum_{s}m_{s}\Omega_{s}=0$.
The formulation of the chemical production rates $\Omega_{s}$ is
taken from \cite{FluctReactFHD} and summarized in Appendix \ref{app:DiffusionMatrix}.

Diffusion is driven by the gradients of the electrochemical potentials
\begin{equation}
\mu_{s}(\V x,T,P)=\mu_{s}^{0}(T,P)+\frac{k_{B}T}{m_{s}}\log(x_{s}\gamma_{s})+z_{s}\Phi,\label{mus}
\end{equation}
where $\mu_{s}^{0}(T,P)$ is a reference chemical potential and $\gamma_{s}(\V x,T,P)$
is the activity coefficient (for an ideal mixture, $\gamma_{s}=1$).
This gives the dissipative diffusive mass fluxes \cite{LowMachElectrolytes}
\begin{equation}
\overline{\M F}=-\rho\M W\M{\chi}\left(\M{\Gamma}\grad{\V x}+\frac{\bar{m}\V W\V z}{k_{B}T}\grad\Phi\right),\label{detmassflux}
\end{equation}
where $\M{\chi}$ is a symmetric positive semi-definite diffusion
matrix that can be computed from the Maxwell-Stefan diffusion coefficients
\cite{LowMachMultispecies,FluctReactFHD}. Here $\M{\Gamma}$ is the
matrix of thermodynamic factors,
\begin{equation}
\M{\Gamma}=\M I+\left(\M X-\V x\V x^{T}\right)\M H,\label{eq:Gamma_F}
\end{equation}
where the symmetric matrix $\M H$ is the Hessian of the excess free
energy per particle; for an ideal mixture $\M H=\M 0$ and $\M{\Gamma}$
is the identity matrix \cite{LowMachMultispecies}. The stochastic
mass fluxes $\widetilde{\M F}$ are given by the fluctuation-dissipation
relation, 
\begin{equation}
\widetilde{\M F}=\sqrt{2\bar{m}\rho}\;\M W\M{\chi}^{\frac{1}{2}}\V{\mathcal{Z}}^{\text{mass}},\label{stochmassflux}
\end{equation}
where $\M{\chi}^{\frac{1}{2}}$ is a ``square root'' of $\M{\chi}$
satisfying $\M{\chi}^{\frac{1}{2}}(\M{\chi}^{\frac{1}{2}})^{T}=\M{\chi}$,
and $\M{\mathcal{Z}}^{\text{mass}}(\V r,t)$ is a standard Gaussian
random vector field with uncorrelated components.

In summary, the composition follows the equation (\ref{eq:ddtw}),
with electrodiffusive fluxes given by the sum of (\ref{detmassflux})
and (\ref{stochmassflux}); the chemical production rates are discussed
in Appendix \ref{app:DiffusionMatrix} and given by (\ref{Rs}).

For dilute species, the expression for the electrodiffusive dissipative
fluxes reduces to that in the familiar PNP equations. Specifically,
for a species $s$ that is dilute, $x_{s}\ll1$, we get the familiar
Nernst-Planck-Fick law (see Appendix A in \cite{FluctReactFHD})
\begin{equation}
\overline{\V F}_{s}\approx-\rho\frac{m_{s}D_{s}}{\bar{m}_{\text{solv}}}\left(\grad x_{s}+\frac{\bar{m}_{\text{solv}}w_{s}z_{s}}{k_{B}T}\grad\Phi\right)=-\rho D_{s}\left(\grad w_{s}+\frac{m_{s}w_{s}z_{s}}{k_{B}T}\grad\Phi\right),\label{Eq:FickDilute}
\end{equation}
where $x_{s}\approx\bar{m}_{\text{solv}}w_{s}/m_{s}$, and $\bar{m}_{\text{solv}}=\left(\sum_{\text{solvent }s^{\prime}}\;w_{s^{\prime}}/m_{s^{\prime}}\right)^{-1}$
is the mixture-averaged molecular mass of the solvent, which could
itself be a mixture of liquids. Here $D_{s}$ is the trace diffusion
coefficient of the dilute species in the solvent, which can be related
to the Maxwell-Stefan coefficients involving species $s$ (see (40)
in \cite{FluctReactFHD}). The stochastic flux also simplifies in
the fluctuating PNP equations for dilute species,
\[
\widetilde{\M F}_{s}\approx\sqrt{2\rho m_{s}w_{s}D_{s}}\,\V{\mathcal{Z}}_{s}^{\text{mass}}.
\]

The boundary conditions for (\ref{eq:ddtw}) depend on the nature
of the physical boundary. We consider \emph{non-reactive} impermeable
\emph{walls} and \emph{reservoirs}; reactive boundaries can be accounted
for \cite{ElectroneutralAsymptotics_Reactions} but we do not consider
them here. For both kinds of boundaries the normal component of the
velocity is zero in the Boussinesq approximation (see Eq. (15) in
\cite{LowMachExplicit} for a generalization to low Mach number variable-density
models). This implies that the normal mass fluxes of all species at
walls must be zero, $\V F^{(n)}=\V F\cdot\V n=\V 0$. Reservoir boundaries
are intended to model a permeable membrane that connects the system
to a large reservoir held at a specified concentration $\V w_{\text{resvr}}$,
and correspond to a Dirichlet condition on $\V w$.

\subsection{\label{subsec:Electroneutral}Electroneutral Formulation}

The charged-fluid equations (\ref{eq:PoissonEq},\ref{ddtvel},\ref{divvel},\ref{eq:ddtw})
suffer from a well-known stiffness: The characteristic Debye length
scale $\lambda_{D}$ is typically much smaller than the macroscopic/device
scales of interest. Thin Debye boundary layers develop near physical
boundaries, with thickness proportional to $\lambda_{D}$. Outside
of these layers, the fields vary much more smoothly on scales much
larger than the Debye length. On such scales, the electrolyte is effectively
\emph{electroneutral}, and electrodiffusion is described the electroneutral
limit of the PNP equations \cite{Electrophysiology_Peskin_Review,ElectroneutralAsymptotics_Yariv}.

The electroneutral bulk equations can be justified by formal asymptotic
analysis \cite{Electrophysiology_Peskin_Review,ElectroneutralAsymptotics_Yariv}.
This analysis leads to an elliptic equation for the potential $\Phi$
that forces the evolution to preserve charge neutrality. Here we derive
this equation by simply invoking charge neutrality as a local linear
\emph{constraint},
\begin{equation}
Z=\rho\V z^{T}\V w=\rho\sum_{s=1}^{N_{s}}z_{s}w_{s}=0,\label{eq:Electroneutrality}
\end{equation}
everywhere in the bulk. By differentiating the constraint $Z\left(\V r,t\right)=0$
we get
\begin{equation}
\frac{\partial Z}{\partial t}=\V z^{T}\frac{\partial}{\partial t}\left(\rho\V w\right)=\sum_{s=1}^{N_{s}}z_{s}\left(-\divg(\rho w_{s}\V v)-\divg\V F_{s}+m_{s}\Omega_{s}\right)=0.\label{eq:dz_dt}
\end{equation}
Because advection preserves $Z=0$,
\begin{equation}
\sum_{s=1}^{N_{s}}z_{s}\divg(\rho w_{s}\V v)=\divg\left(\left(\rho\sum_{s=1}^{N_{s}}z_{s}w_{s}\right)\V v\right)=\divg\left(Z\V v\right)=0,\label{eq:advection_eln}
\end{equation}
and reactions conserve charge, $\sum_{s=1}^{N_{s}}z_{s}m_{s}\Omega_{s}=0$,
Eq. (\ref{eq:dz_dt}) simplifies to
\begin{equation}
\sum_{s=1}^{N_{s}}z_{s}\divg\V F_{s}=\divg\left(\sum_{s=1}^{N_{s}}z_{s}\V F_{s}\right)=\divg\left(\V z^{T}\V F\right)=\V 0.\label{eq:div_Jz_zero}
\end{equation}

\subsubsection{Electroneutral Elliptic Equation}

Using the expressions (\ref{detmassflux}) and (\ref{stochmassflux})
for the diffusive mass fluxes, we can rewrite the condition $\divg\left(\V z^{T}\V F\right)=\V 0$
as an elliptic PDE for the electric potential,
\begin{equation}
\divg\left[\left(\frac{\bar{m}\rho}{k_{B}T}\V z^{T}\V W\V{\chi}\V W\V z\right)\grad\Phi\right]=\divg\left(\V z^{T}\V F_{\mathrm{d}}\right),\label{eq:ElectroneutralPoisson}
\end{equation}
where $\V F_{\mathrm{d}}$ denotes the diffusive fluxes without the
electrodiffusion,
\[
\V F_{\mathrm{d}}=-\rho\M W\M{\chi}\M{\Gamma}\grad{\V x}+\sqrt{2\bar{m}\rho}\;\M W\M{\chi}^{\frac{1}{2}}\V{\mathcal{Z}}^{\text{mass}}.
\]
We see that in the electroneutral limit, the electric potential becomes
a Lagrange multiplier that enforces the electroneutrality condition.
It is given by the solution of the modified elliptic equation (\ref{eq:ElectroneutralPoisson}),
and \emph{not} by the quasielectrostatic Poisson equation (\ref{eq:PoissonEq}).
In summary, the fluctuating electroneutral equations we consider in
this work are given by (\ref{ddtvel},\ref{divvel},\ref{eq:ddtw},\ref{eq:ElectroneutralPoisson}).

It is worth pointing out that the validity of the electroneutral limit
requires that $\lambda_{D}$ be small \emph{everywhere} in the bulk,
where we recall that for an ideal solution $\lambda_{D}\sim\left(\V z^{T}\M W\M M\V z\right)^{-1/2}$.
This requires the presence of some charges everywhere in the domain,
that is, one cannot use (\ref{eq:ElectroneutralPoisson}) when parts
of the domain are ion-free since in those parts of the domain $\lambda_{D}$
would diverge; an example of a situation not covered by the electroneutral
limit would be the diffusive mixing of pure and salty water. In particular,
(\ref{eq:ElectroneutralPoisson}) is not uniformly elliptic if in
some part of the domain $\V z^{T}\V W\V{\chi}\V W\V z\rightarrow0$,
i.e., if $z_{s}w_{s}\rightarrow0$ for all species $s$. In practice,
for water solutions, it is energetically very unfavorable to remove
all ions and purify water to the point where the Debye length would
approach macroscopic/device scales\footnote{For ultra-pure water the ion mass fractions are $\approx10^{^{-10}}$and
the Debye length is a few microns.}. We will therefore assume here that there are sufficiently many ions
everywhere in the domain to justify the electroneutral limit (\ref{ddtvel},\ref{divvel},\ref{eq:ddtw},\ref{eq:ElectroneutralPoisson}).

\subsubsection{Boundary Conditions}

Obtaining proper boundary conditions for the electroneutral equations
(\ref{ddtvel},\ref{divvel},\ref{eq:ddtw},\ref{eq:ElectroneutralPoisson})
requires a nontrivial asymptotic analysis matching the electroneutral
bulk ``outer solution'' on the outside of the Debye layer to the
boundary layer ``inner solution'' inside the Debye layer \cite{ElectroneutralAsymptotics_Yariv,ElectroneutralAsymptotics_Dielectric,ElectroneutralAsymptotics_Metals}.
Since we are interested here in the electroneutral bulk, what we mean
by boundary conditions are the conditions not on the physical boundary
itself but rather on the outer boundary of the Debye layer. In the
electroneutral limit $\lambda_{D}/l_{\min}\rightarrow0$, where $l_{\min}$
is the smallest length scale of interest, the thickness of the boundary
layer is formally zero and the outer conditions become effective boundary
conditions for the electroneutral bulk equations. Though surface reactions
can affect the charge density bound to dielectric boundaries (e.g.,
electron exchange), in this paper we do not consider surface chemistry.

Here we will assume that there is \emph{no surface conduction} in
the Debye layer, i.e., we only need to consider normal mass fluxes
(the curved surface analysis in \cite{ElectroneutralAsymptotics_Yariv}
shows that curvature does not enter in the leading-order asymptotics)
at the outer edge of the double layer. For dielectric boundaries,
a careful analysis of the validity of the assumption of no surface
currents is carried out in \cite{ElectroneutralAsymptotics_Dielectric},
and it is concluded that it is valid only for weakly to moderately
charged surfaces. For highly charged dielectric boundaries, surface
conduction enters even in the leading-order asymptotic matching. For
metals, a careful asymptotic analysis is carried out in \cite{ElectroneutralAsymptotics_Metals}
and shows that, in regions where the potential jump across the layer
is exponentially large (measured with respect to the thermal voltage
$k_{B}T/e$), surface conduction also enters.

Under the assumption of no surface conduction, we first consider the
boundary conditions for the electrodiffusion equations (\ref{eq:ddtw},\ref{eq:ElectroneutralPoisson}),
and then turn our attention to the velocity equations (\ref{ddtvel},\ref{divvel}).
We recall that the electrodiffusive mass flux is
\[
\V F=\V F_{\mathrm{d}}-\left(\frac{\bar{m}\rho}{k_{B}T}\V W\V{\chi}\V W\V z\right)\grad\Phi.
\]
Since the flux must \emph{locally} preserve the charge neutrality,
$\V z^{T}\V F^{(n)}=0$ on the boundary, where we recall that $\V F^{(n)}=\V n\cdot\V F$
denotes the fluxes normal to the boundary. This immediately gives
the effective Neumann boundary condition for the potential,
\begin{equation}
\frac{\partial\Phi}{\partial n}=\left(\frac{\bar{m}\rho}{k_{B}T}\V z^{T}\V W\V{\chi}\V W\V z\right)^{-1}\,\left(\V z^{T}\V F_{\mathrm{d}}^{(n)}\right),\label{eq:phi_BC}
\end{equation}
where $\V F_{\mathrm{d}}^{(n)}=\V n\cdot\V F_{\mathrm{d}}$.

For dilute solutions and impermeable walls, in the deterministic
case, one can show that $\V F_{\mathrm{d}}^{(n)}=\V 0$ which means
that (\ref{eq:phi_BC}) becomes a \emph{homogeneous} Neumann condition
for the potential, $\partial\Phi/\partial n=0$, which is the boundary
condition for a dielectric boundary with no bound surface charge in
the charged-fluid formulation. The derivation is well-known for binary
dilute electrolytes \cite{ElectroneutralAsymptotics_Yariv}, and it
is straightforward to generalize it to dilute multi-ion solutions
as follows. From the electroneutrality condition $\V z^{T}\V w=0$
we get $\V z^{T}\left(\partial\V w/\partial n\right)=0$. Focusing
on the dilute ions only, we have from (\ref{Eq:FickDilute}) that
the vanishing of the electrodiffusive flux is equivalent to
\[
\frac{\partial\V w}{\partial n}+\frac{\left(\partial\Phi/\partial n\right)}{k_{B}T}\M M\V W\V z=0.
\]
Taking the dot product with $\V z$ of both sides of this equation,
we obtain $\partial\Phi/\partial n=0$ because $\V z^{T}\M M\V W\V z\sim\lambda_{D}^{-2}>0$.
This implies $\partial\V w/\partial n=0$ and therefore $\V F_{\mathrm{d}}^{(n)}=\V 0$.
We will assume that $\partial\Phi/\partial n=0$ also holds on impermeable
walls for general mixtures and even in the presence of fluctuations,
even though we have not been able to rigorously justify this. It is
worthwhile to note that \emph{any} choice of Neumann boundary condition
on the potential gives identical total electrodiffusive flux $\V F^{(n)}$;
the only physically-relevant boundary condition is that $\V F^{(n)}$
vanish at impermeable walls. Only the Lorentz force in (\ref{ddtvel}),
which depends on $\Phi$, is affected by the choice of Neumann boundary
condition. While the Lorentz force is essential for modeling electrokinetic
flows, it plays a minimal role in the problems we study here, so our
choice to enforce a homogeneous condition $\partial\Phi/\partial n=0$
for impermeable walls is inconsequential.

For reservoir boundaries, $\V F_{\mathrm{d}}^{(n)}$ is known at the
boundary from the Dirichlet conditions on $\V w$, and (\ref{eq:phi_BC})
becomes an \emph{inhomogeneous} Neumann condition for the potential.
In summary, at a physical boundary, we impose the following boundary
conditions for (\ref{eq:ddtw},\ref{eq:ElectroneutralPoisson}):
\begin{itemize}
\item $\V F_{\mathrm{d}}^{(n)}=\V n\cdot\V F_{\mathrm{d}}=\V 0$ and $\partial\Phi/\partial n=0$
for impermeable walls.
\item $\V w=\V w_{\text{resvr}}$ with $\V z^{T}\mathbf{w}_{\mathrm{resvr}}=0$
and (\ref{eq:phi_BC}) for reservoirs.
\end{itemize}
Note that the condition (\ref{eq:phi_BC}) applies irrespective of
whether the boundary (wall or membrane) is dielectric (polarizable)
or metal (conducting); the effective condition for the potential is
always Neumann, even if in the charged-fluid formulation there is
a Dirichlet condition on the potential.

The electroneutral boundary conditions for the velocity equation (\ref{ddtvel},\ref{divvel})
are even harder to derive. In general, the fluid velocity on the outer
boundary of the Debye layer is \emph{not} zero, even for a no-slip
boundary. This means that the appropriate velocity boundary condition
for the electroneutral equations is a \emph{specified-slip} condition,
$v_{n}=0$ and $\V v_{\tau}$ nonzero. However, to our knowledge,
the velocity slip has only been computed using asymptotic analysis
for binary electrolytes, and this analysis has not, to our knowledge,
been generalized to multi-ion mixtures. For dilute electrolytes, slip
expressions have been proposed without a careful asymptotic analysis,
see for example (4) in \cite{MultiIonDiffusiophoresis}. Because
an asymptotic analysis for multispecies electrolytes is not available,
and because in the example we consider here there are no applied electric
fields or highly-charged surfaces (either of which could make the
apparent slip velocity large enough to play some role), in this work
we will simply use the \emph{same} boundary condition (no-slip or
free-slip) for the electroneutral and the charged-fluid formulations.
For no-slip boundaries this means $\V v_{\tau}=\V 0$, which is expected
to be a good approximation for dielectric boundaries if the surface
charge density is sufficiently small. We emphasize, however, that
an effective no-slip boundary condition is \emph{not} appropriate
in general (e.g., slip is important for ionic diffusiophoresis), and
each specific application requires a careful consideration of the
boundary condition.

\subsubsection{\label{subsubsec:AmbipolarDiffusion}Effective Salt Diffusion}

Let us define the vector field
\begin{equation}
\V g_{\text{amb}}^{(\Phi)}=\left(\frac{\bar{m}\rho}{k_{B}T}\V z^{T}\V W\V{\chi}\V W\V z\right)^{-1}\V z^{T}\V F_{\mathrm{d}},\label{eq:gradphi_amb}
\end{equation}
which simplifies for deterministic models of dilute electrolytes to
\begin{equation}
\V g_{\text{amb}}^{(\Phi)}=-\left(k_{B}T\right)\frac{\sum_{s}z_{s}D_{s}\grad w_{s}}{\sum_{s}m_{s}z_{s}^{2}D_{s}w_{s}}.\label{eq:grad_phi_approx}
\end{equation}
For some specific special cases, the solution of the effective Poisson
equation (\ref{eq:ElectroneutralPoisson}) is given \footnote{Note that the boundary condition (\ref{eq:phi_BC}) is consistent
with $\grad\Phi=\V g_{\text{amb}}^{(\Phi)}$ on the boundary (cf.
(\ref{eq:gradphi_amb})).} by $\grad\Phi=\V g_{\text{amb}}^{(\Phi)}$. This is sometimes stated
as a fact, see for example Eq. (3) in \cite{MultiIonDiffusiophoresis}.
However, except for dilute binary electrolytes and for problems in
one dimension, $\V g_{\text{amb}}^{(\Phi)}$ is, in general, not a
gradient, unless $\V w$ (and thus the denominator in (\ref{eq:grad_phi_approx}))
can be approximated to be constant over the domain.

For dilute \emph{binary} electroneutral electrolytes $w_{2}=-z_{1}w_{1}/z_{2}$,
which in the absence of fluctuations gives
\begin{equation}
\left(k_{B}T\right)^{-1}\V g_{\text{amb}}^{(\Phi)}=-\frac{\left(D_{1}-D_{2}\right)}{\left(m_{1}z_{1}D_{1}-m_{2}z_{2}D_{2}\right)}\frac{\grad w_{1}}{w_{1}}=-\frac{\left(D_{1}-D_{2}\right)}{\left(m_{1}z_{1}D_{1}-m_{2}z_{2}D_{2}\right)}\grad\left(\ln\,w_{1}\right),\label{eq:gradphi_amb_binary}
\end{equation}
which is indeed a gradient of a function and therefore $\grad\Phi=\V g_{\text{amb}}^{(\Phi)}$.
Substituting (\ref{eq:gradphi_amb_binary}) into Fick's law (\ref{Eq:FickDilute})
we obtain 
\begin{equation}
\overline{\V F}_{1}=-\rho\frac{\left(m_{1}z_{1}-m_{2}z_{2}\right)D_{1}D_{2}}{\left(m_{1}z_{1}D_{1}-m_{2}z_{2}D_{2}\right)}\grad w_{1}.\label{eq:FickAmbipolar}
\end{equation}
The effective diffusion coefficient of the salt is therefore the weighted
harmonic mean
\begin{equation}
D^{\text{amb}}=\frac{\left(m_{1}z_{1}-m_{2}z_{2}\right)D_{1}D_{2}}{\left(m_{1}z_{1}D_{1}-m_{2}z_{2}D_{2}\right)},\label{eq:D_amb}
\end{equation}
which we will refer to as the \emph{ambipolar} binary diffusion coefficient.

This shows that for a dilute binary electrolyte \emph{without} fluctuations
the electroneutral model is equivalent to modeling a binary salt as
an uncharged substance with an effective ambipolar diffusion coefficient,
i.e., the two ions tend to diffuse together. However, this correspondence
is not true in more general cases. Specifically, it is not valid when
fluctuations are included, when the system is not dilute, or when
there are more than two ions. Although the equivalence is lost, a
number of prior studies \cite{InstabilityRT_Chemistry,DiffusiveInstability_AcidBase,FluctReactFHD}
have used (\ref{eq:D_amb}) to define effective diffusion coefficients
of salts in more general situations. We will refer to this type of
approach as the \emph{ambipolar approximation} and investigate it
in detail in Section \ref{subsec:Fingering}.

\subsection{\label{subsec:Fluctuations}Thermal Fluctuations}

Important quantities that can be derived from the fluctuating hydrodynamics
equations are the spectra of the fluctuations at thermodynamic equilibrium,
referred to as the static structure factors. These structure factors
can be obtained from either the more general results derived in \cite{LowMachElectrolytes}
for the charged-fluid equations, or, from equilibrium statistical
mechanics. It is important to confirm that our electroneutral FHD
equations give the correct spectrum of fluctuations in order to justify
our formulation of the stochastic fluxes.

The matrix of equilibrium structure factors can be expressed either
in terms of mass or mole fractions. Here we define the matrix of static
structure factors in terms of the fluctuations in the mass fractions
$\d{\V w}$ around the equilibrium concentrations, which for notational
brevity we denote in this section with $\V w$ without any decoration,
\begin{equation}
S_{s,s^{\prime}}\left(\V k\right)=\av{\left(\widehat{\d w}_{s}\left(\V k\right)\right)\left(\widehat{\d w}_{s^{\prime}}\left(\V k\right)\right)^{\star}}.\label{eq:S_k}
\end{equation}
where $s$ and $s^{\prime}$ are two species (including $s=s^{\prime}$),
$\V k$ is the wavevector, hat denotes a Fourier transform, and star
denotes a complex conjugate.

The static factors for an electrolyte mixture with an arbitrary number
of species at thermodynamic equilibrium are \cite{LowMachMultispecies,LowMachElectrolytes}
\begin{equation}
\M S=\M S_{0}-\frac{1}{\left(k^{2}\lambda_{D}^{2}+1\right)}\frac{\M S_{0}\V z\V z^{T}\M S_{0}}{\V z^{T}\M S_{0}\V z},\label{S_k_eq_general}
\end{equation}
where the structure factor for a mixture of uncharged species (i.e.,
for $\V z=0$) is \cite{LowMachMultispecies} 
\begin{equation}
\M S_{0}=\frac{\bar{m}}{\rho}\left(\M W-\V w\V w^{T}\right)\left[\M{\Gamma}\left(\M X-\V x\V x^{T}\right)+\V 1\V 1^{T}\right]^{-1}\left(\M W-\V w\V w^{T}\right),
\end{equation}
and $\V 1$ is the vector of 1's. The Debye length can be generalized
to non-ideal mixtures as
\begin{equation}
\lambda_{D}^{-2}=\frac{\rho^{2}}{\epsilon k_{B}T}\V z^{T}\M S_{0}\V z=\frac{\bar{m}\rho}{\epsilon k_{B}T}\V z^{T}\M W\left[\M{\Gamma}\left(\M X-\V x\V x^{T}\right)+\V 1\V 1^{T}\right]^{-1}\M W\V z.\label{Debye_general}
\end{equation}
See Eqs. (41) and (42) in \cite{LowMachMultispecies} for a simplification
for ideal mixtures, including dilute solutions.

The structure factor $S_{Z}\left(\V k\right)$ of the total free charge
density $Z=\rho\V z^{T}\V w$ is
\begin{equation}
S_{Z}=\rho^{2}\left\langle \left(\V z^{T}\widehat{\delta\V w}\right)\left(\V z^{T}\widehat{\delta\V w}\right)^{\star}\right\rangle =\rho^{2}\V z^{T}\M S\V z.
\end{equation}
Using the generalized definition of the Debye length (\ref{Debye_general})
allows to conveniently express it as 
\begin{equation}
S_{Z}=\rho^{2}\V z^{T}\M S\V z=\left(\rho^{2}\V z^{T}\M S_{0}\V z\right)\frac{k^{2}\lambda_{D}^{2}}{1+k^{2}\lambda_{D}^{2}}.\label{S_z}
\end{equation}
The fact that $S_{Z}(\V k)$ tends to zero for small wavenumbers ($k\lambda_{D}\rightarrow0$)
is a manifestation of the transition to the electroneutral regime
at large length scales. It is important to point out that at scales
much larger than the Debye length the fluctuations $\d{\V w}$ are
electroneutral in addition to the electroneutral mean composition
$\V w$. This means that the composition $\V w+\d{\V w}$ \emph{strictly}
remains on the electroneutral constraint at all times, consistent
with our electroneutral formulation.

At thermodynamic equilibrium, for length scales much larger than the
Debye length, the structure factor (\ref{S_k_eq_general}) simplifies
to
\begin{equation}
\M S^{(\text{eln})}=\lim_{k\lambda_{D}\rightarrow0}\M S=\M S_{0}-\frac{\M S_{0}\V z\V z^{T}\M S_{0}}{\V z^{T}\M S_{0}\V z},\label{S_k_eq_eln}
\end{equation}
and this is the spectrum of fluctuations in composition in the electroneutral
limit. Note that $\M S^{(\text{eln})}\V z=\V 0$, as expected from
the electroneutrality. For dilute electrolyte solutions, which are
necessarily ideal, if we consider two ions of species $s$ and $s^{\prime}$,
we have the explicit formula
\begin{equation}
S_{s,s^{\prime}}^{(\text{eln})}=\rho^{-1}\left(w_{s}m_{s}\delta_{s,s^{\prime}}-\left(\sum_{k}m_{k}z_{k}^{2}w_{k}\right)^{-1}\left(w_{s}m_{s}z_{s}\right)\left(w_{s^{\prime}}m_{s^{\prime}}z_{s^{\prime}}\right)\right).\label{eq:S_ideal_eln}
\end{equation}
In Appendix \ref{app:EntropyArgument} we derive the same result for
a dilute binary electrolyte using equilibrium statistical mechanics,
without referring to the generalized PNP equations.

It can be confirmed that the electroneutral equations (\ref{ddtvel},\ref{divvel},\ref{eq:ddtw},\ref{eq:ElectroneutralPoisson})
are consistent with (\ref{S_k_eq_eln}), which demonstrates that nothing
special needs to be done in the electroneutral limit to handle the
fluctuating diffusive fluxes except to include them in the right-hand
side of the modified Poisson equation (\ref{eq:ElectroneutralPoisson}).

\section{\label{sec:NumericalAlgorithm}Numerical Algorithm}

In this section we describe our charged-fluid and electroneutral numerical
algorithms, both of which are second-order accurate in space and time
deterministically, and second-order weakly accurate for the linearized
fluctuating hydrodynamics equations. The algorithms are closely based
on the algorithm developed for isothermal constant-density reactive
multispecies mixtures of non-ionic species in our prior work \cite{FluctReactFHD}.
The handling of the charged species and in particular the quasielectrostatic
Poisson equation is described already in detail in our prior work
\cite{LowMachElectrolytes}. Here we only briefly sketch the algorithmic
details and focus on the key differences with our prior work.

We note that special care is taken to ensure that the \emph{only}
difference between the charged-fluid and electroneutral algorithms
is that a different elliptic equation is solved to compute the electric
potential $\Phi$. We therefore present both cases together and note
any differences explicitly where necessary.

\subsection{Spatial Discretization}

Our spatial discretization of reaction-advection-diffusion is identical
to the one used in our previous work \cite{FluctReactFHD}, which
is itself a slight modification of the methods described in \cite{LowMachImplicit,LowMachMultispecies,LowMachElectrolytes},
improved to accurately handle very small numbers of molecules. The
spatial discretization is based on a structured-grid finite-volume
approach with cell-averaged densities, electric potential, and pressure,
and face-averaged (staggered) velocities. We use standard second-order
stencils for the gradient, divergence, and spatial averaging in order
to satisfy discrete fluctuation-dissipation balance (DFDB) \cite{LLNS_S_k}.

The discretization of the momentum equation (\ref{ddtvel},\ref{divvel}),
including no-slip or free-slip boundary conditions, is the same as
our previous works \cite{LowMachImplicit,LowMachMultispecies,LowMachElectrolytes,FluctReactFHD},
with the important caveat that the Lorentz force is evaluated as $\grad\cdot\left(\epsilon\grad\Phi\right)\grad\Phi$
so that the same implementation works for either the charged-fluid
or the electroneutral formulations. Standard centered second-order
stencils are used to discretize $\grad\cdot\left(\epsilon\grad\Phi\right)\grad\Phi$
on the faces of the grid.

The discretization of the electrodiffusion equations (\ref{eq:ddtw},\ref{eq:ElectroneutralPoisson})
is closely based on that developed in \cite{LowMachElectrolytes,FluctReactFHD}.
Our implementation independently tracks the densities of all species
$\rho_{s}$ but ensures the overall mass conservation in the Boussinesq
limit, $\sum_{s=1}^{N_{s}}\rho_{s}=\rho_{0}$, in each grid cell to
within (Stokes) solver tolerance. For each species, we construct the
mass fluxes on faces of the grid and employ the standard conservative
divergence in order to guarantee conservation of mass for each species.
Diffusive fluxes, including the dissipative and stochastic fluxes,
are computed as described in \cite{FluctReactFHD}. Chemical reaction
terms are local and are computed independently in each cell as in
\cite{FluctReactFHD}.

The elliptic equations (\ref{eq:PoissonEq}) and (\ref{eq:ElectroneutralPoisson})
are discretized using a standard centered second-order stencil, and
the resulting linear system is solved using a geometric multigrid
algorithm \cite{LowMachElectrolytes}. For the electroneutral elliptic
equation (\ref{eq:ElectroneutralPoisson}), $\V W$ and $\V W\V{\chi}$
are already computed on each grid face to calculate diffusive mass
fluxes (see \cite{FluctReactFHD} for details), and therefore the
non-constant coefficient $\sim\V z^{T}\V W\V{\chi}\V W\V z$ can be
directly computed on each grid face.

Boundary conditions for the electroneutral electrodiffusion equations
(\ref{eq:ddtw},\ref{eq:ElectroneutralPoisson}) are implemented as
follows. For impermeable walls, the condition $\V F_{\mathrm{d}}^{(n)}=\V n\cdot\V F_{\mathrm{d}}=\V 0$
is trivially implemented in our finite-volume scheme by zeroing the
total mass flux (including the stochastic fluxes) on the boundary.
The modified elliptic equation (\ref{eq:ElectroneutralPoisson}) is
then solved with the homogeneous Neumann condition $\partial\Phi/\partial n=0$.
For reservoirs, the Dirichlet condition $\V w=\V w_{\text{resvr}}$
is implemented by computing $\grad\V x$ at the boundary using one-sided
differences and the specified values on the boundary; this then gives
the dissipative portion of the diffusive mass flux $\overline{\M F}_{\mathrm{d}}=-\rho\M W\M{\chi}\M{\Gamma}\grad{\V x}$.
The generation of the stochastic component of the diffusive mass flux
$\widetilde{\M F}$ at the boundary is described in prior work \cite{LLNS_Staggered}.
Once $\V F_{\mathrm{d}}=\overline{\M F}_{\mathrm{d}}+\widetilde{\M F}$
is computed on the boundary, (\ref{eq:ElectroneutralPoisson}) is
solved with an inhomogeneous Neumann condition computed using (\ref{eq:phi_BC}).

Advective mass fluxes $\V{\rho}_{f}\V v$ are computed on each face
$f$ of the grid by first computing face-centered densities $\V{\rho}_{f}=\rho\V w_{f}=\rho_{0}\V w_{f}$.
Our implementation supports two ways to compute face-centered densities
$\V{\rho}_{f}$. \emph{Centered advection} uses two-point averaging
of densities to faces, and is non-dissipative and thus preserves DFDB
\cite{LLNS_S_k}. However, in order to prevent nonphysical oscillations
in mass densities in high Péclet number flows with sharp gradients,
we also use the Bell-Dawson-Shubin (BDS) second-order \emph{Godunov
advection} scheme \cite{BDS}; more details about how the BDS scheme
is used in our numerical implementations can be found in \cite{LowMachImplicit}.
We note that BDS advection adds artificial dissipation and does not
obey a fluctuation-dissipation principle, but is necessary for simulations
where centered advection would fail due to insufficient spatial resolution.

An additional complication that arises in the electroneutral limit
is ensuring that advection preserves electroneutrality, i.e., ensuring
that the spatial discretization maintains the continuum identity (\ref{eq:advection_eln}).
Since the advection velocity used in our discretization is discretely
divergence free, advection automatically maintains linear constraints
on the cell-centered densities \emph{if} the face densities $\V{\rho}_{f}$
satisfy the linear constraint for each face $f$. For centered advection
this is automatic because the face densities are computed by linear
interpolation. For BDS advection, however, the face-centered densities
are computed using a complicated space-time extrapolation that involves
nonlinear ``limiters,'' and are not guaranteed to satisfy the same
linear constraints as the cell-centered densities. It is therefore
necessary to project the densities back onto all linear constraints.
One such constraint is the mass conservation $\sum_{s}\rho_{s}=\rho_{0}$
and the other is the electroneutrality $\sum_{s}z_{s}\rho_{s}=0$.
Assume we are given a composition $\V w_{f}=\V{\rho}_{f}/\rho_{0}$
on face $f$, which does not necessarily satisfy the two constraints
$\V 1^{T}\V w=1$ and $\V z^{T}\V w=0$. The projection onto \emph{both}
constraints can be accomplished with the following sequence of updates:
\begin{align*}
\V w_{f} & \leftarrow\V w_{f}-\frac{\V z^{T}\V w_{f}}{\V z^{T}\V z}\V z,\\
\V w_{f} & \leftarrow\frac{\V w_{f}}{\V 1^{T}\V w_{f}}.
\end{align*}
The first update is a standard $L^{2}$ projection onto the plane
$\V z^{T}\V w=0$, and the second is a simple rescaling that preserves
$\V 1^{T}\V w=1$. This choice of projections is not unique and does
not affect the second-order accuracy for smooth problems.

\subsection{Temporal Discretization}

Our second-order temporal integrator is taken from our prior work
\cite{FluctReactFHD} and is summarized in Section \ref{subsec:Algorithm}.
Unlike the trapezoidal predictor-corrector used in \cite{LowMachElectrolytes},
here we use the \emph{midpoint} predictor-corrector method described
in \cite{FluctReactFHD} to accommodate our treatment of chemical
reactions, and to dramatically improve the robustness for large Schmidt
number \cite{FluctReactFHD}. Furthermore, the Boussinesq approximation
allows us to simplify the algorithm compared to the low Mach version
presented in \cite{LowMachElectrolytes}. Note that relative to the
algorithm in \cite{FluctReactFHD} we need to precompute some terms
related to charged species; however, the overall update strategy remains
the same. In particular, in the absence of charged species our algorithm
is equivalent to that presented in \cite{FluctReactFHD}.

Our algorithm introduces an important correction term in the right-hand
side (r.h.s.) in the modified elliptic equation (\ref{eq:ElectroneutralPoisson})
in the corrector step. Namely, numerical tests revealed that errors
due to finite tolerances in the iterative elliptic solver lead to
a slow drift away from electroneutrality over many time steps. This
drift can be prevented by modifying the elliptic equation as follows.
Consider an Euler update of the form
\[
\frac{\V w\left(t+\D t\right)-\V w\left(t\right)}{\D t}=-\grad\cdot\left(\V F_{\mathrm{d}}-\left(\frac{\bar{m}\rho}{k_{B}T}\V W\V{\chi}\V W\V z\right)\grad\Phi\right).
\]
Requiring electroneutrality at the end of the step, $\V z^{T}\V w\left(t+\D t\right)=0$,
without assuming electroneutrality at the beginning of the step, we
obtain the corrected elliptic equation at time $t$,
\begin{equation}
\divg\left[\left(\frac{\bar{m}\rho}{k_{B}T}\V z^{T}\V W\V{\chi}\V W\V z\right)\grad\Phi\right]=\divg\left(\V z^{T}\V F_{\mathrm{d}}\right)-\left(\frac{\V z^{T}\V w}{\D t}\right).\label{eq:ElectroneutralPoisson_num}
\end{equation}
In our numerical algorithm, we only employ this correction to the
elliptic equation in the corrector step. We have found this to be
sufficient and to lead to a stable algorithm that maintains the charge
neutrality to a relative error below solver tolerances \footnote{We have found that in some cases (e.g., equal diffusion coefficients
for all species) the r.h.s. of (\ref{eq:ElectroneutralPoisson_num})
is analytically (nearly) zero so that numerically it is dominated
by numerical noise. This make the elliptic solver do unnecessary work
if we use a standard relative error tolerance based on the magnitude
of the r.h.s. Instead, we use a tolerance $\delta\divg\left(\abs{\V z}^{T}\V F_{\mathrm{d}}\right)$
based on the absolute values of the charges per mass, where $\delta\sim10^{-12}-10^{-9}$
is a relative tolerance for the iterative solver.}. In practice, we find that the numerical errors introduced by the
iterative geometric multigrid elliptic solvers create localized charges
but not a global charge; therefore the spatial average of the r.h.s.
of (\ref{eq:ElectroneutralPoisson_num}) is zero within roundoff tolerance.
Nevertheless, in order to ensure that the elliptic equation (\ref{eq:ElectroneutralPoisson_num})
is solvable, in our implementation we subtract from the r.h.s. it's
spatial average.

By adding charges to the 3-species mixture test described in Section
III.C.1 of \cite{LowMachMultispecies}, we have verified (not shown)
that our algorithm/code reproduces the correct spectrum of electroneutral
equilibrium fluctuations (\ref{S_k_eq_eln}), for both ideal and non-ideal
mixtures, for either periodic, reservoir, or impermeable boundaries.
This validates our formulation and implementation of the stochastic
mass flux (including boundary conditions). We have also verified (not
shown) second-order deterministic accuracy for the acid-base fingering
example by initializing the simulations from a smoothly perturbed
sine-wave interface.

\subsection{\label{subsec:Algorithm}Summary of Algorithm}

We now summarize the $n^{th}$ time step that computes state at time
$t^{n+1}=\left(n+1\right)\D t$ from the state at time $t^{n}=n\D t$.
Superscripts denote the time point at which certain quantities are
evaluated, for example, $\V f^{n+\myhalf,*}=\V f\left(\V w^{n+\myhalf,*},\,\left(n+1/2\right)\D t\right)$
denotes the buoyancy force estimated at the midpoint. We denote with
$\left(\M{\mathcal{W}}^{\text{mom}}\right)^{n}$ and $\left(\V{\mathcal{W}}_{(1)}^{\text{mass}}\right)^{n}$
(for the predictor stage) and $\left(\V{\mathcal{W}}_{(2)}^{\text{mass}}\right)^{n}$(for
the corrector stage) collections of i.i.d.\ (independent and identically
distributed) standard normal random variables generated on control
volume faces independently at each time step, and $\overline{\M{\mathcal{W}}}^{\text{mom}}\equiv\M{\mathcal{W}}^{\text{mom}}+\big(\M{\mathcal{W}}^{\text{mom}}\big)^{T}$.
We denote collections of independent Poisson random variables generated
at cell centers independently at each time step with $\mathcal{P}_{(1)}$
(predictor stage) and $\mathcal{P}_{(2)}$ (corrector stage), and
denote $[\bullet]^{+}\equiv\max(\bullet,0)$. The notation for computing
the divergence of the advective fluxes using the BDS scheme is defined
and explained in Section III.B.1 in \cite{LowMachImplicit}. We remind
the reader that $\rho=\rho_{0}$ is a constant in the Boussinesq approximation,
maintained by our code to roundoff tolerance.

The $n^{th}$ predictor-corrector update consists of the following
steps:
\begin{enumerate}
\item Calculate predictor diffusive fluxes (deterministic and stochastic),
\begin{equation}
\V F_{\mathrm{d}}^{n}=\left(-\rho\M W\V{\chi}\V{\Gamma}\grad\V x\right)^{n}+\sqrt{\frac{2\bar{m}\rho}{\D V\D t/2}}\left(\M W\V{\chi}^{\frac{1}{2}}\right)^{n}\left(\V{\mathcal{W}}_{(1)}^{\text{mass}}\right)^{n}.
\end{equation}
\item Solve the predictor elliptic equation for $\Phi^{n}$,
\begin{equation}
\begin{cases}
\divg\left(\epsilon^{n}\grad\Phi^{n}\right)=-Z^{n}\text{ if charged-fluid, otherwise} & \text{}\\
\divg\left[\left(\frac{\bar{m}\rho}{k_{B}T}\V z^{T}\V W\V{\chi}\V W\V z\right)^{n}\grad\Phi^{n}\right]=\divg\left(\V z^{T}\V F_{\mathrm{d}}^{n}\right).
\end{cases}
\end{equation}
\item Calculate predictor electrodiffusive fluxes $\V F^{n}$ and chemical
production rates $\V R^{n}$,
\begin{align}
\V F^{n}= & \V F_{\mathrm{d}}^{n}-\left(\frac{\bar{m}\rho}{k_{B}T}\M W\V{\chi}\M W\V z\right)^{n}\grad\Phi^{n},\\
R_{s}^{n}= & \frac{1}{\D V\D t/2}\sum_{r}\sum_{\alpha=\pm}\D{\nu}_{sr}^{\alpha}\mathcal{P}_{(1)}\left(\left(a_{r}^{\alpha}\right)^{n}\D V\D t/2\right).
\end{align}
\item Solve the predictor Stokes system for $\V v^{n+1,*}$ and $\pi^{n+\myhalf,*}$:
$\divg\V v^{n+1,*}=0$ and
\begin{align*}
\frac{\left(\rho\V v\right)^{n+1,*}-\left(\rho\V v\right)^{n}}{\D t}+\grad\pi^{n+\myhalf,*} & =\divg\left(\rho\V v\V v^{T}\right)^{n}+\frac{1}{2}\divg\left(\eta^{n}\overline{\grad}\V v^{n}+\eta^{n}\overline{\grad}\V v^{n+1,*}\right)+\V f^{n}\\
 & +\divg\left(\sqrt{\frac{\eta^{n}k_{B}T}{\D V\D t}}\left(\overline{\M{\mathcal{W}}}^{\text{mom}}\right)^{n}\right)+\big[\left(\divg(\epsilon\grad\Phi)\right)\grad\Phi\big]^{n}.
\end{align*}
\item Calculate predictor mass densities,
\begin{equation}
\rho_{s}^{n+\myhalf,*}=\rho_{s}^{n}+\frac{\D t}{2}\left[-\divg\V F_{s}^{n}+m_{s}R_{s}^{n}\right]-\frac{\D t}{2}\divg\begin{cases}
\rho_{s}^{n}\left(\frac{\V v^{n}+\V v^{n+1,*}}{2}\right)\text{ if centered}\\
\textrm{BDS}\left(\rho_{s}^{n},\frac{\V v^{n}+\V v^{n+1,*}}{2},\divg\V F_{s}^{n},\frac{\Delta t}{2}\right).
\end{cases}
\end{equation}
\item Calculate corrector diffusive fluxes,
\begin{equation}
\V F_{\mathrm{d}}^{n+\myhalf,*}=\left(-\rho\M W\V{\chi}\V{\Gamma}\grad\V x\right)^{n+\myhalf,*}+\sqrt{\frac{2\bar{m}\rho}{\D V\D t/2}}\left(\M W\V{\chi}^{\frac{1}{2}}\right)^{n+\myhalf,*}\left(\frac{\left(\V{\mathcal{W}}_{(1)}^{\text{mass}}\right)^{n}+\left(\V{\mathcal{W}}_{(2)}^{\text{mass}}\right)^{n}}{\sqrt{2}}\right).
\end{equation}
\item Solve the corrector elliptic equation for $\Phi^{n+\myhalf,*}$,
\begin{equation}
\begin{cases}
\divg\left(\epsilon^{n+\myhalf,*}\grad\Phi^{n+\myhalf,*}\right)=-Z^{n+\myhalf,*}\text{ if charged-fluid, otherwise} & \text{}\\
\divg\left[\left(\frac{\bar{m}\rho}{k_{B}T}\V z^{T}\V W\V{\chi}\V W\V z\right)^{n+\myhalf,*}\grad\Phi^{n+\myhalf,*}\right]=\divg\left(\V z^{T}\V F_{\mathrm{d}}^{n+\myhalf,*}\right)-\D t^{-1}\left(\V z^{T}\V w^{n+\myhalf,*}\right). & \text{}
\end{cases}
\end{equation}
\item Calculate corrector diffusive fluxes $\V F^{n+\myhalf,*}$ and chemical
production rates $\V R^{n+\myhalf,*}$,
\begin{align}
\V F^{n+\myhalf,*}= & \V F_{\mathrm{d}}^{n+\myhalf,*}-\left(\frac{\bar{m}\rho}{k_{B}T}\M W\V{\chi}\M W\V z\right)^{n+\myhalf,*}\grad\Phi^{n+\myhalf,*},\\
R_{s}^{n+\myhalf,*}= & \frac{1}{2}\left[R_{s}^{n}+\frac{1}{\D V\D t/2}\sum_{r}\sum_{\alpha=\pm}\D{\nu}_{sr}^{\alpha}\mathcal{P}_{(2)}\left(\left(2(a_{r}^{\alpha})^{n+\myhalf,*}-(a_{r}^{\alpha})^{n}\right)^{+}\D V\D t/2\right)\right].
\end{align}
\item Update the mass densities,
\begin{equation}
\rho_{s}^{n+1}=\rho_{s}^{n}+\D t\left[-\divg\V F_{s}^{n+\myhalf,*}+m_{s}R_{s}^{n+\myhalf,*}\right]-\D t\divg\begin{cases}
\rho_{s}^{n+\myhalf,*}\left(\frac{\V v^{n}+\V v^{n+1,*}}{2}\right)\text{ if centered}\\
\textrm{BDS}\left(\rho_{s}^{n},\frac{\V v^{n}+\V v^{n+1,*}}{2},\divg\V F_{s}^{n+\myhalf,*},\Delta t\right).
\end{cases}
\end{equation}
\item Solve the corrector Stokes systems for $\V v^{n+1}$ and $\pi^{n+\myhalf}$:
$\divg\V v^{n+1}=0$ and
\begin{align*}
\frac{\left(\rho\V v\right)^{n+1}-\left(\rho\V v\right)^{n}}{\D t}+\grad\pi^{n+\myhalf} & =-\frac{1}{2}\divg\left[\left(\rho\V v\V v^{T}\right)^{n}+\left(\rho\V v\V v^{T}\right)^{n+1,*}\right]+\frac{1}{2}\divg\left(\eta^{n}\bar{\grad}\V v^{n}+\eta^{n+1}\bar{\grad}\V v^{n+1}\right)\\
 & +\frac{1}{2}\divg\left[\left(\sqrt{\frac{\eta^{n}k_{B}T}{\D V\D t}}+\sqrt{\frac{\eta^{n+1}k_{B}T}{\D V\D t}}\right)\left(\overline{\M{\mathcal{W}}}^{\text{mom}}\right)^{n}\right]+\V f^{n+\myhalf,*}\\
 & +\big[\left(\divg(\epsilon\grad\Phi)\right)\grad\Phi\big]^{n+\myhalf,*}.
\end{align*}
\end{enumerate}

\section{\label{sec:AcidBase}Numerical Study of Acid-Base Neutralization}

In our previous work \cite{FluctReactFHD} we studied the development
of asymmetric fingering patterns arising from a gravitational instability
in the presence of a neutralization reaction. In particular, we performed
the first three-dimensional simulations of a double-diffusive instability
occurring during the mixing of dilute aqueous solutions of HCl and
NaOH in a vertical Hele-Shaw cell, as studied experimentally in \cite{DiffusiveInstability_AcidBase}.
In this prior study, as in all other theoretical and computational
studies of this kind of instability \cite{InstabilityRT_Chemistry,DiffusiveInstability_AcidBase},
we treated HCl, NaOH, and NaCl as uncharged species in the spirit
of the ambipolar approximation described in Section \ref{subsubsec:AmbipolarDiffusion}.
The acid-base neutralization reaction was written as
\begin{equation}
\mathrm{HCl}+\mathrm{NaOH}\rightarrow\mathrm{NaCl}+\mathrm{H_{2}O}.\label{neutralization_rxn}
\end{equation}

In reality, however, the acid, the base, and the salt are all strong
electrolytes and essentially completely disassociate into $\mathrm{Na}^{+}$,
$\mathrm{Cl}^{-}$, $\mathrm{H}^{+}$, and $\mathrm{OH}^{-}$ ions,
and the neutralization reaction is simply the (essentially irreversible)
formation of water,
\[
\mathrm{H}^{+}+\mathrm{OH}^{-}\rightarrow\mathrm{H_{2}O},
\]
with $\mathrm{Na}^{+}$ and $\mathrm{Cl}^{-}$ being spectator ions.
An important feature of this system is that the trace diffusion coefficients
of the four ions are very different; specifically, using cgs units
($\text{cm}^{2}/\text{s}$) the literature values are $D_{\mathrm{{Na^{+}}}}=1.33\cdot10^{-5}$,
$D_{\mathrm{{Cl^{-}}}}=2.03\cdot10^{-5}$, $D_{\mathrm{{H^{+}}}}=9.35\cdot10^{-5}$,
and $D_{\mathrm{{OH^{-}}}}=5.33\cdot10^{-5}$. Although the ambipolar
approximation is only strictly valid for dilute binary systems, we
define effective diffusion coefficients for the neutral species by
harmonic averages of the anion and cation diffusion coefficients following
(\ref{eq:FickAmbipolar}) to obtain: $D_{\mathrm{{HCl}}}=3.34\cdot10^{-5}$,
$D_{\mathrm{{NaOH}}}=2.13\cdot10^{-5}$, and $D_{\mathrm{{NaCl}}}=1.61\cdot10^{-5}$.

Simulating this instability using the charged-fluid formulation would
be infeasible because the length scales of interest are on the millimeter
scale. In this work we use the electroneutral formulation to study
this instability and assess the (in)accuracy of the commonly-used
ambipolar approximation. Numerical studies based on the ambipolar
approximation showed that the fingering instability can be triggered
on a realistic time scale (as compared to experiments) purely by thermal
fluctuations, without any artificial perturbations of the initial
interface \cite{FluctReactFHD}. The studies also demonstrated that
the effect of fluctuations is dominated by the contribution of the
stochastic \emph{momentum} flux, and not by fluctuations in the initial
condition, the stochastic mass flux, or the stochastic chemical production
rate. This can be understood as a consequence of the fact that advection
by thermal velocity fluctuations, which are driven by the stochastic
momentum flux, leads to \emph{giant concentration fluctuations} in
the presence of sharp concentration gradients \cite{FluctHydroNonEq_Book,GiantFluctuations_Theory,FractalDiffusion_Microgravity}.
These nonequilibrium fluctuations completely dominate equilibrium
fluctuations at the scales of interest, and are sufficiently large
to drive the fingering instability. We have confirmed that the same
conclusions apply when charges are accounted for. Therefore, in the
simulations reported here we do \emph{not} include stochastic mass
fluxes and reaction rate fluctuations, and initialize the simulations
from a deterministic initial condition with a sharp interface between
the acid and the base and zero fluid velocity.

Because of the importance of giant concentration fluctuations to the
formation of the instability, in Section \ref{subsec:GiantFluct}
we first validate our algorithm and implementation by computing the
spectrum of giant fluctuations in a ternary electrolyte. Then, we
study the fingering instability at an acid-base neutralization front
in Section \ref{subsec:Fingering}.

\subsection{\label{subsec:GiantFluct}Giant Nonequilibrium Fluctuations in Electroneutral
Ternary Mixtures}

In this section we examine the giant concentration fluctuations in
a non-reactive dilute electroneutral ternary electrolyte in the presence
of a steady applied concentration gradient. Giant fluctuations in
a binary electrolyte were studied using the charged-fluid formulation
in Section V.B.2 in \cite{LowMachElectrolytes}. It was concluded
there that for small wavenumbers, $k\lambda_{D}\ll1$, the electroneutral
nonequilibrium fluctuations in a binary electrolyte can be described
using the ambipolar formulation, as expected. Specifically, the spectrum
of the giant fluctuations is the same as it would be in a solution
with a single neutral species diffusing with the ambipolar diffusion
coefficient (\ref{eq:FickAmbipolar}). However, this conclusion no
longer holds for solutions with three or more charged species, even
if dilute.

Therefore, here we examine the spectrum of the giant fluctuations
in a dilute solution of three ions with valencies $V_{1}=V_{2}=1$
and $V_{3}=-1$, in the absence of gravity or reactions. In order
to focus on the nonequilibrium fluctuations we omit the stochastic
mass flux from (\ref{eq:ddtw}), so that the fluctuations are generated
entirely by the random velocity. In arbitrary units in which $k_{B}=1$
and $e=1$, we set $\rho=1$, $T=1$ and assume equal molecular masses
$m_{1}=m_{2}=m_{3}=1$, and trace diffusion coefficients $D_{1}=1$,
$D_{2}=1/2$, and $D_{3}=3/2$. The viscosity is set to $\eta=10^{3}$
to give a realistically large Schmidt number $\mathrm{Sc}\sim10^{3}$,
and we set $\epsilon=0$, which makes $\lambda_{D}=0$ and removes
the (fluctuating) Lorentz force from the momentum equation. The domain
is quasi-two dimensional with $L_{x}=L_{y}=64$ discretized with $64\times64$
cells with grid spacing $\D x=\D y=1$. The thickness of the domain
is set to $\D z=10^{6}$ to give weak fluctuations that can be described
by the linearized fluctuating hydrodynamics equations. We impose equal
and opposite macroscopic gradients for the co-ion species and no gradient
for the counter-ion using reservoir boundary conditions, with imposed
$w_{1}=4.5\cdot10^{-3}$, $w_{2}=5.5\cdot10^{-3}$, and $w_{3}=10^{-2}$
at the $y=0$ boundary, and $w_{1}=5.5\cdot10^{-3}$, $w_{2}=4.5\cdot10^{-3}$,
and $w_{3}=10^{-2}$ at the $y=L_{y}$ boundary. We set the time step
size to $\D t=0.05$ and perform a total of $10^{6}$ time steps skipping
$10^{5}$ steps in the beginning to allow the system to reach the
steady state, after which we collect statistics on the spectrum of
fluctuations $\M S\left(k_{x},k_{y}=0\right)$.

\begin{figure}
\centering{}\includegraphics[width=0.9\textwidth]{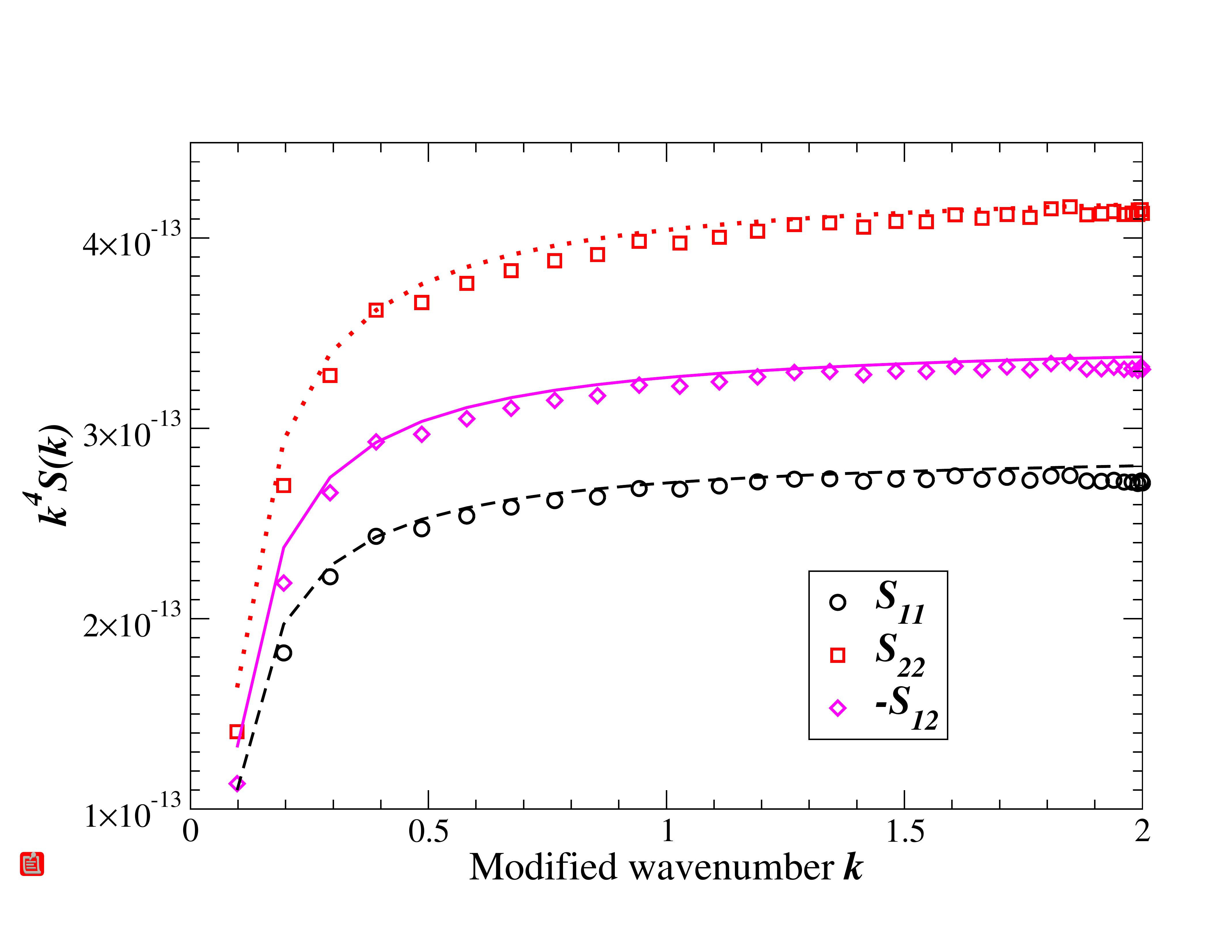}\caption{\label{fig:S_k_giant_ternary}Spectrum of giant nonequilibrium concentration
fluctuations in a ternary electrolyte mixture in the presence of an
imposed concentration gradient of the two co-ions. To account for
errors in the discrete approximation to the continuum Laplacian, the
$x$ axis shows the modified wavenumber $\tilde{k}_{x}=\sin(k_{x}\protect\D x/2)/\left(\protect\D x/2\right)$
instead of $k_{x}$. Numerical results (symbols) for the components
of $\tilde{k}_{x}^{4}\,\protect\M S\left(k_{x},k_{y}=0\right)$ corresponding
to the co-ions (positive $S_{11}$ and $S_{22}$, and negative $S_{12}$)
are compared to the theoretical prediction (\ref{eq:S_k_giant_ternary})
(lines).}
\end{figure}

The theoretical spectrum of the giant fluctuations for a dilute ternary
electrolyte can be computed by following the computation described
in Section III.C in \cite{LowMachElectrolytes}, and then taking the
electroneutral limit $k\lambda_{D}\rightarrow0$. The same result
can also be obtained from the electroneutral equations directly; with
the help of a computer algebra system the limit $k\lambda_{D}\rightarrow0$
of the charged-fluid $\M S\left(\V k\right)$ is straightforward to
compute, so we follow that route. Just as for non-ionic solutions,
a $k_{x}^{-4}$ power law is observed until the confinement effect
becomes significant for small $k_{x}\ll L_{y}^{-1}$. Following \cite{LowMachElectrolytes},
we multiply the theoretical result for an unconfined bulk system by
a confinement factor \cite{GiantFluctFiniteEffects_NoGravity} to
obtain
\begin{equation}
S_{s,s^{\prime}}\left(k_{x},k_{y}=0\right)=f_{ss^{\prime}}\,\frac{k_{B}T}{\eta D_{1}}\,\frac{1}{k_{x}^{4}}\,\left[1+\frac{4\left(1-\cosh(k_{x}L_{y})\right)}{k_{x}L_{y}\left(k_{x}L_{y}+\sinh(k_{x}L_{y})\right)}\right],\label{eq:S_k_giant_ternary}
\end{equation}
where our theoretical calculations predict $f_{11}=147/124$, $f_{22}=219/124$,
and $f_{12}=-177/124$. Note that it is sufficient to examine only
the part of $\M S$ corresponding to two of the charged ions (here
the two co-ions), since electroneutrality dictates the spectra involving
the third ion, and conservation of mass dictates the spectra involving
the solvent. In Fig. \ref{fig:S_k_giant_ternary} we compare our numerical
results to the theoretical predictions (\ref{eq:S_k_giant_ternary}),
and find good agreement for all three structure factors.

\subsection{\label{subsec:Fingering}Fingering Instability at an Acid-Base Front}

In this section we investigate the development of asymmetric fingering
patterns arising from a diffusion-driven gravitational instability
in the presence of a neutralization reaction. This system has been
studied experimentally and theoretically using a two-dimensional Darcy
advection-diffusion-reaction model \cite{InstabilityRT_Chemistry,DiffusiveInstability_AcidBase}
based on an ambipolar approximation where the acid, base, and salt
are treated as uncharged species instead of as disassociated ions.
Here we perform large-scale three-dimensional simulations with the
ions treated as individual species.

Since the fingering instability is driven by the small changes of
density with composition, it is crucial to first match the dependence
of density on composition between the ambipolar diffusion (molecule-based)
model used in Section IV.D in \cite{FluctReactFHD} and the electrodiffusion
(ion-based) model used in this work. Following \cite{InstabilityRT_Chemistry,DiffusiveInstability_AcidBase,FluctReactFHD},
we assume that the solution density is linearly dependent on the solute
concentrations in both cases, which is reasonable since the solutions
are dilute. This gives the buoyancy force
\begin{equation}
\V f(\V w)=-\rho\left(\sum_{\mathrm{{solute}\:}s}\frac{\alpha_{s}}{M_{s}}w_{s}\right)g\V e_{y},\label{eq:buyoancy_force}
\end{equation}
where $\alpha_{s}$ is the solutal expansion coefficient, $M_{s}$
is the molecular weight (in g/mol) of solute $s$, and the gravitational
acceleration $\V g=-g\V e_{y}$ acts in the negative $y$ direction.
For the ambipolar model, the values of $\alpha_{s}$ for $s=\mathrm{{HCl}},\:\mathrm{{NaOH}},\:\mathrm{{NaCl}}$
are obtained from Table~II in \cite{DiffusiveInstability_AcidBase}.
For the ionic model, we compute the four unknown coefficients $\alpha_{s}$
for $s=\mathrm{{Na^{+}}},\:\mathrm{{Cl^{-}}},\:\mathrm{{H^{+}}},\:\mathrm{{OH^{-}}}$
by matching the dependence of density on composition between the two
models for electroneutral binary solutions of HCl, NaOH, and NaCl.
Only three independent coefficients $\alpha_{s}$ matter because electroneutrality
fixes the concentration of the fourth ion, so we arbitrarily require
that $\mathrm{Na}^{+}$ and $\mathrm{Cl}^{-}$ have the same coefficient
$\alpha_{\mathrm{{Na^{+}}}}=\alpha_{\mathrm{{Cl^{-}}}}$. It is important
to observe that this procedure matches the density between an arbitrary
dilute solution of HCl, NaOH, and NaCl and the corresponding ionic
solution resulting after the molecules disassociate completely into
$\mathrm{Na}^{+}$, $\mathrm{Cl}^{-}$, $\mathrm{H}^{+}$, and $\mathrm{OH}^{-}$
ions. The reverse is not possible, that is, one cannot take an arbitrary
solution of the ions and uniquely determine a corresponding molecular
solution. In particular, a solution of only $\mathrm{H}^{+}$ and
$\mathrm{OH}^{-}$ would not have a physically-reasonable density
according to our model. We will validate shortly that any differences
we see between the molecular and ionic models of the instability stem
from the difference between standard Fickian diffusion and electrodiffusion,
and not from our procedure for matching the buoyancy force.

For the model setup and physical parameters, we follow Section IV.D
in \cite{FluctReactFHD} and mimic the experiment of Lemaigre et al.~\cite{DiffusiveInstability_AcidBase}.
We use cgs units unless otherwise specified and assume $T=293$ and
atmospheric pressure, neglecting any heat release in the reaction
as justified in \cite{InstabilityRT_Chemistry}. We set $g=981$,
$\rho=1$, and $\eta=0.01$. We consider a Hele-Shaw cell with side
lengths $L_{x}=L_{y}=1.6$ and $L_{z}=0.05$, with the $y$-axis pointing
in the vertical direction, and the $z$-axis being perpendicular to
the glass plates. The domain is divided into $512\times512\times16$
grid cells, which is twice finer than the grid used in \cite{FluctReactFHD}
in order to better-resolve the sharp interface in the early stages
of the mixing. We start with a gravitationally stable initial configuration,
where an aqueous solution of NaOH with molarity 0.4~mol/L is placed
on top of a denser aqueous solution of HCl with molarity 1~mol/L.
We impose periodic boundary conditions in the $x$ direction, no-slip
impermeable walls in the $z$ direction, and in the $y$ direction
we use free-slip reservoir boundary conditions with imposed concentrations
that match the initial conditions of each layer. We use BDS advection
because of the presence of an initially sharp interface.

Since the neutralization equilibrium lies far to the product side,
we only consider the forward reaction. We use the law of mass action
for a dilute solution (\ref{eq:LMA_n}), and express the reaction
propensity in terms of number densities, $a^{+}=k\,n_{\mathrm{{HCl}}}\,n_{\mathrm{{NaOH}}}$
for the molecule-based model, and $a^{+}=k\,n_{\mathrm{{H^{+}}}}\,n_{\mathrm{{OH^{-}}}}$
for the ion-based model \footnote{Observe that the reaction rates are matched between the molecule-based
model and the corresponding ion-based model because the number density
of HCl/NaOH in the non-ionic mixture matches the number density of
$\mathrm{H}^{+}$/$\mathrm{OH}^{-}$ in the corresponding ionic mixture.}. In reality, neutralization is a diffusion-limited reaction that
occurs extremely fast (with rate $\lambda\sim10^{11}\:\mathrm{s}^{-1}$),
essentially as soon as reactants encounter each other. The estimated
value of $k\sim10^{-11}\:\mathrm{cm^{3}s^{-1}}$ is impractically
large, and would require an unreasonably small grid spacing to resolve
the penetration depth (which would be on molecular scales), and an
unreasonably small time step size to resolve the reactions. For our
simulations, we choose a smaller value $k=10^{-19}$ that is an order
of magnitude smaller than the one used in \cite{FluctReactFHD}, and
enlarge the time step size to $\D t=10^{-2}$ by an order of magnitude
accordingly. Deterministic numerical studies presented in Appendix
B in \cite{FluctReactFHD} show that increasing the rate beyond $k=10^{-19}-10^{-18}$
hardly changes the results, so we believe our simulation parameters
are realistic. Nevertheless, our main goal here is to compare molecule-
and ion-based models and assess the accuracy of the ambipolar approximation,
so in this study it is more important to resolve the spatio-temporal
scales in the problem than to match experimental observations.

\begin{figure}
\centering{}\includegraphics[width=0.99\textwidth]{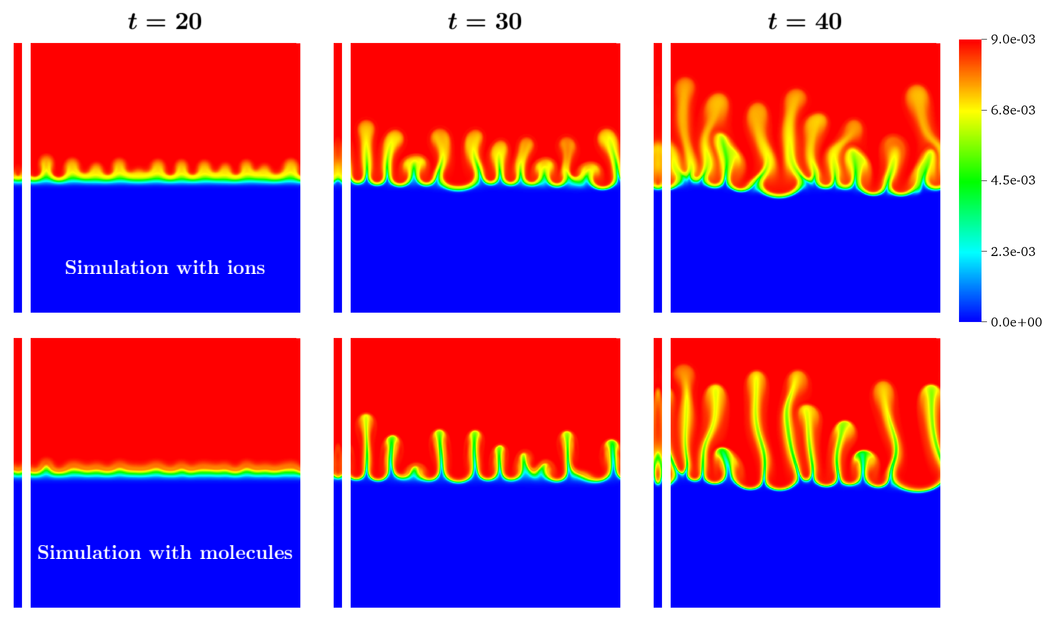}\caption{\label{fig:AcidBaseReal}Asymmetric growth of convective fingering
patterns in a Hele-Shaw cell, induced by a gravitational instability
in the presence of a neutralization reaction. The top row corresponds
to electroneutral electrodiffusion of ions, while the bottom row corresponds
to ambipolar diffusion of acid, base, and salt molecules; both simulations
use the same random numbers for the stochastic momentum flux. The
density of $\mathrm{Na}^{+}$ is shown with a color scale at 20, 30,
and 40 seconds (columns going left to right) from the beginning of
the simulation, initialized without any fluctuations. Two-dimensional
slices of the three-dimensional field $\rho_{\mathrm{{Na^{+}}}}(x,y,z)$
are shown. The square images show $\rho_{\mathrm{{Na^{+}}}}(x,y,z=L_{z}/2)$
(halfway between the glass plates) and the thin vertical side bars
show the slice $\rho_{\mathrm{{Na^{+}}}}(x=0,y,z)$ corresponding
to the left edge of the square images.}
\end{figure}

\begin{figure}
\centering{}\includegraphics[width=0.99\textwidth]{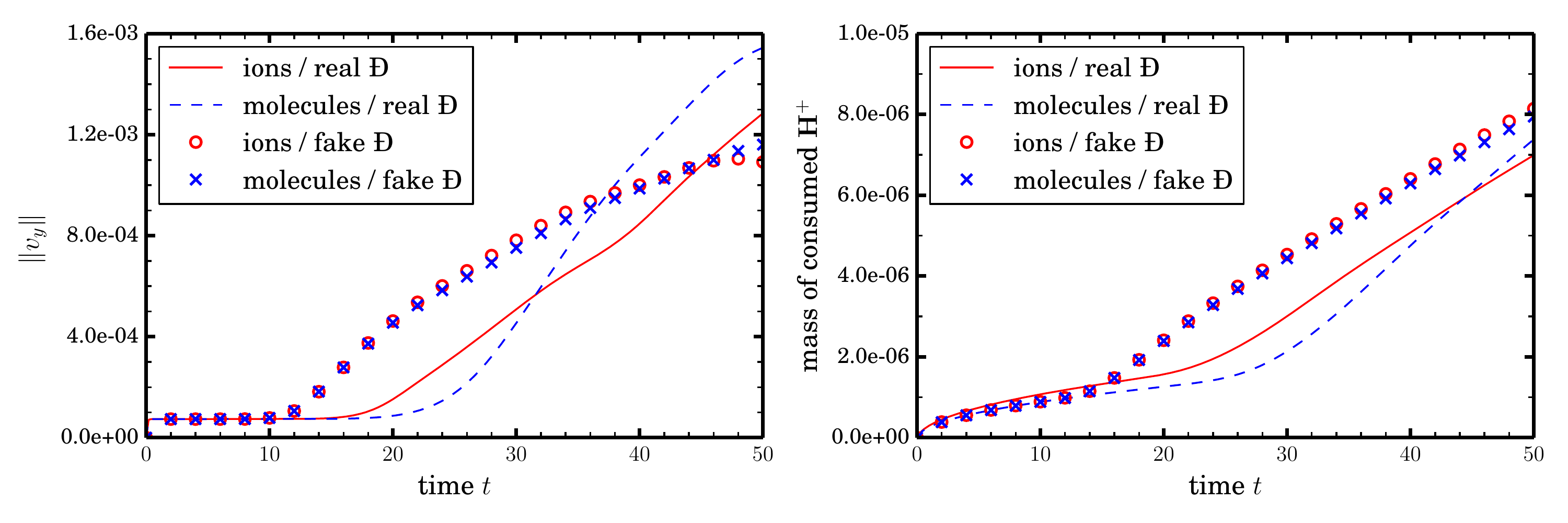}\caption{\label{fig:AcidBaseComparison}Time dependence of the norm of vertical
velocity (left panel) and of the total mass of $\mathrm{H}^{+}$ consumed
in the neutralization reaction (right panel). We compare simulations
where the species are ions versus those where the species are neutral
molecules (see legend). Lines are results based on the true tabulated
diffusion values for the ions, while symbols show results for fake
values of the ion diffusion coefficients, artificially made to be
closer to each other.}
\end{figure}

In Fig. \ref{fig:AcidBaseReal} we compare the density profiles of
$\mathrm{Na}^{+}$ between the model based on electroneutral electrodiffusion
with ions, and that based on molecules using ambipolar diffusion coefficients.
For the molecule-based simulations, we compute the density of $\mathrm{Na}^{+}$
assuming that the acid is completely disassociated. To enable a direct
comparison between the two cases, we employ the same sequence of pseudorandom
numbers for the stochastic momentum flux in both cases. Although the
development of the instability follows similar trends in the two cases,
there are clearly-visible differences between the top and bottom rows
in the figure. For example, the $\mathrm{Na}^{+}$ fingers develop
sooner and diffuse more for the ion-based simulations. These differences
can also be seen by comparing the lines in Fig. \ref{fig:AcidBaseComparison},
where we show the norm of the $y$ component of velocity (corresponding
to the progress of the instability) and the total mass of consumed
$\mathrm{H}^{+}$ (corresponding to the production of salt in the
molecule-based model) as a function of time. Our findings clearly
demonstrate that \emph{quantitative} predictions can only be made
by solving the complete electroneutral electrodiffusion equations
presented here. The ambipolar approximation can only be used as a
\emph{qualitative} model of the instability.

\begin{figure}
\centering{}\includegraphics[width=0.99\textwidth]{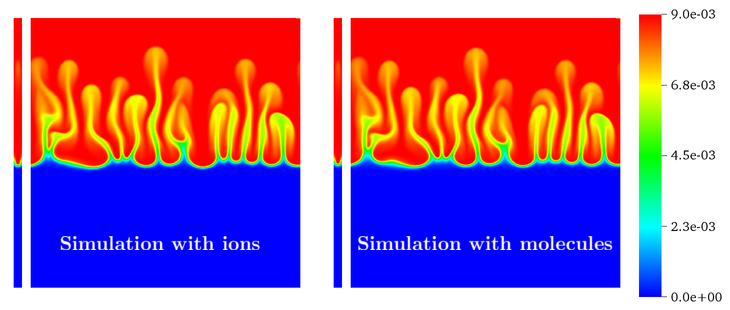}\caption{\label{fig:AcidBaseFake}Density of $\mathrm{Na}^{+}$ at time $t=30$s
for the ion-based model (left panel) or the molecule-based model (right
panel); both simulations use the same random numbers for the stochastic
momentum flux. These simulations use fake values of the ion diffusion
coefficients, artificially made to be closer to each other in order
to make the difference between electrodiffusion and ambipolar diffusion
smaller.}
\end{figure}

To demonstrate that the clear difference between electrodiffusion
and ambipolar diffusion is caused by the large difference in the true
diffusion coefficients of the ions ($D_{\mathrm{{Na^{+}}}}=1.33\cdot10^{-5}$,
$D_{\mathrm{{OH^{-}}}}=5.33\cdot10^{-5}$, $D_{\mathrm{{H^{+}}}}=9.35\cdot10^{-5}$,
and $D_{\mathrm{{Cl^{-}}}}=2.03\cdot10^{-5}$), we also perform simulations
where we artificially match the diffusion coefficients for the reactant
ions and molecules by setting them to fake values, $D_{\mathrm{{Na^{+}}}}=D_{\mathrm{{OH^{-}}}}=D_{\mathrm{{NaOH}}}=2.13\cdot10^{-5}$
(i.e., $\mathrm{Na}^{+}$ and $\mathrm{OH}^{-}$ ions diffuse with
the same coefficient as NaOH) and $D_{\mathrm{{H^{+}}}}=D_{\mathrm{{Cl^{-}}}}=D_{\mathrm{{HCl}}}=3.34\cdot10^{-5}$
(i.e., $\mathrm{H}^{+}$ and $\mathrm{Cl}^{-}$ ions diffuse with
the same coefficient as HCl). In the molecule-based simulations, the
diffusion coefficient of NaCl is set to the harmonic average of the
two fake diffusion coefficients of $\mathrm{Na}^{+}$ and $\mathrm{Cl}^{-}$
ions, $D_{\mathrm{{NaCl}}}=2.6\cdot10^{-5}$. Figure \ref{fig:AcidBaseFake}
compares the density of $\mathrm{Na}^{+}$ in ion- and molecule-based
simulations using these artificial (fake) values of the diffusion
coefficients. The two panels in Fig. \ref{fig:AcidBaseFake} are visually
almost indistinguishable, showing very little difference between electrodiffusion
and ambipolar diffusion, unlike the panels in Fig. \ref{fig:AcidBaseReal}.
This is further demonstrated by the symbols in Figure \ref{fig:AcidBaseComparison}.
This demonstrates that the difference between electrodiffusion and
standard Fickian diffusion is large when the multiple ions involved
diffuse with widely varying coefficients.

\section{\label{sec:Conclusions}Conclusions}

We formulated the electroneutral reactive generalized PNP equations
and included thermal fluctuations using fluctuating hydrodynamics
and the chemical master equation. The only difference between the
charged-fluid equations and their electroneutral limit is in the elliptic
equation for the electric potential. We presented a second-order midpoint
predictor-corrector scheme for both sets of equations. We studied
giant nonequilibrium fluctuations in ternary electrolytes in the electroneutral
limit, and demonstrated that our numerical algorithm accurately reproduces
theoretical predictions. We also modeled a fingering instability at
an acid-base mixing front and demonstrated that modeling the acid,
base, and salt as neutral species diffusing with ambipolar diffusion
coefficients leads to quantitatively-incorrect results unless the
diffusion coefficients of the ions are very similar.

The temporal discretization we used in this work treats mass diffusion
explicitly. It can be shown that the electroneutral integrator used
here is the limit $\D t\gg\lambda_{D}^{2}/D$ of a method for the
charged-fluid equations in which only the potential is treated implicitly,
i.e., the Poisson equation (\ref{eq:PoissonEq}) is imposed at the
end instead of the beginning of an Euler update. A major challenge
for the future is to develop algorithms that treat electroneutral
electrodiffusion implicitly. This would require solving a coupled
linear system for both the composition \emph{and} the electric potential
at the end of the time step. This is in some ways similar to our treatment
of the velocity equation where we solve a Stokes problems for both
velocity and pressure. The main challenge in developing implicit electrodiffusion
discretizations is the development of effective preconditioners for
the coupled electrodiffusion system.

In this work we used Neumann boundary conditions for the potential
that were consistent with electroneutrality under the assumption of
no surface conduction. Future work should carefully derive appropriate
boundary conditions for the electroneutral electrodiffusion equations
using asymptotic analysis, at least in the deterministic context.
In this work we used the same velocity boundary conditions for the
charged-fluid and electroneutral formulations because of the absence
of any asymptotic theory for the effective slip for multispecies mixtures.
It is important to carry out such asymptotic theory, even if only
for the case of small zeta or applied potentials/fields. Finally,
allowing for surface reactions in the formulation also requires changing
the boundary conditions. Future developments in these directions would
allow us to model catalytic micropumps \cite{CatalyticMicroPump}
without having to resolve the thin Debye layers around the catalytic
surfaces.
\begin{acknowledgments}
We thank Jean-Philippe Péraud for help with comparisons between charged-fluid
and electroneutral formulations. We would like to thank Anne De Wit
for helpful discussions regarding gravitational instabilities in the
presence of neutralization reactions, and thank Ehud Yariv for generously
sharing his knowledge about the electroneutral limit. We also acknowledge
informative discussions with Charles Peskin. This material is based
upon work supported by the U.S. Department of Energy, Office of Science,
Office of Advanced Scientific Computing Research, Applied Mathematics
Program under Award Number DE-SC0008271 and under contract No. DE-AC02-05CH11231.
This research used resources of the National Energy Research Scientific
Computing Center, a DOE Office of Science User Facility supported
by the Office of Science of the U.S. Department of Energy under Contract
No. DE-AC02-05CH11231. A. Donev was supported in part by the Division
of Chemical, Bioengineering, Environmental and Transport Systems of
the National Science Foundation under award CBET-1804940.
\end{acknowledgments}

\appendix

\section*{Appendix}

\section{\label{app:DiffusionMatrix}Chemical Production Rates}

In this Appendix we summarize how we compute the (deterministic or
stochastic) chemical production rates $\Omega_{s}$. We consider a
liquid mixture consisting of $N_{s}$ species undergoing $N_{r}$
elementary reversible reactions of the form 
\begin{equation}
\sum_{s=1}^{N_{s}}\nu_{sr}^{+}\mathfrak{M}_{s}\rightleftharpoons\sum_{s=1}^{N_{s}}\nu_{sr}^{-}\mathfrak{M}_{s}\quad(r=1,\dots,N_{r}),
\end{equation}
where $\nu_{sr}^{\pm}$ are molecule numbers, and $\mathfrak{M}_{s}$
are chemical symbols. We define the stoichiometric coefficient of
species $s$ in the forward reaction $r$ as $\D{\nu}_{sr}^{+}=\nu_{sr}^{-}-\nu_{sr}^{+}$
and the coefficient in the reverse reaction as $\D{\nu}_{sr}^{-}=\nu_{sr}^{+}-\nu_{sr}^{-}$.
We assume that each reaction $r$ conserves mass, $\sum_{s=1}^{N_{s}}\D{\nu}_{sr}^{\pm}m_{s}=0$,
and charge , $\sum_{s=1}^{N_{s}}\D{\nu}_{sr}^{\pm}m_{s}z_{s}=0$,
which is suitable for bulk reactions in liquids (we do not consider
surface reactions here). It is important to note that all reactions
must be reversible for thermodynamic consistency, although in practice
some reactions can be effectively considered to be irreversible sufficiently
far from thermodynamic equilibrium.

The mean number of reaction occurrences in a locally well-mixed reactive
cell of volume $\D V$ during an infinitesimal time interval $dt$
is given as $a_{r}^{\pm}\D Vdt$, where $a_{r}^{\pm}$ are the propensity
density functions for the forward/reverse ($+/-$) rates of reaction
$r$. Accordingly, the mean production rate of species $s$ in the
deterministic equations is given as 
\begin{equation}
\overline{\Omega}_{s}=\sum_{r=1}^{N_{r}}\sum_{\alpha=\pm}\D{\nu}_{sr}^{\alpha}a_{r}^{\alpha}.
\end{equation}
The propensity density functions are given by the Law of Mass Action
(LMA) kinetics, suitably generalized to non-dilute mixtures \cite{FluctReactFHD},
\begin{equation}
a_{r}^{\pm}=\kappa_{r}^{\pm}\prod_{s=1}^{N_{s}}\left(x_{s}\gamma_{s}\right)^{\nu_{sr}^{\pm}},\label{eq:LMA_x}
\end{equation}
where $\kappa_{r}^{\pm}(T,P)$ is the rate of the forward/reverse
reaction $r$, and $\gamma_{s}(\V x,T,P)$ is the activity coefficient
of species $s$ (for an ideal mixture, $\gamma_{s}=1$). It is important
to note that propensity density functions (\ref{eq:LMA_x}) are expressed
in terms of \emph{mole fractions} $x_{s}$ (for ideal mixtures) or
activities $x_{s}\gamma_{s}$, and \emph{not} in terms of number densities.
For reactions in a dilute solution (which is necessarily an ideal
solution for sufficiently dilution), mole fractions and number densities
are proportional, $x_{s}\approx\bar{m}_{\text{solv}}n_{s}/\rho$,
and one can alternatively write the LMA in the form
\begin{equation}
a_{r}^{\pm}=k_{r}^{\pm}\prod_{s=1}^{N_{s}}n_{s}^{\nu_{sr}^{\pm}}\text{ for dilute solutions}.\label{eq:LMA_n}
\end{equation}

Following \cite{FluctReactDiff,FluctReactFHD}, we use the Chemical
Master Equation (CME) to describe fluctuations in the reaction rates
for small numbers of reactive molecules. For reactions in a closed
well-mixed cell of volume $\D V$, the change in the number of molecules
$N_{s}$ of species $s$ in a given cell during an infinitesimal time
interval $dt$ is expressed in terms of the number of occurrences
$\mathcal{P}(a_{r}^{\pm}\D Vdt)$ of each reaction $r$, 
\begin{equation}
\Omega_{s}\D V\,dt=\sum_{r=1}^{N_{r}}\sum_{\alpha=\pm}\D{\nu}_{sr}^{\alpha}{\mathcal{P}(a_{r}^{\alpha}\D Vdt)},\label{Rs}
\end{equation}
where $\mathcal{P}(m)$ denotes a Poisson random variable with mean
$m$. Note that the instantaneous rate of change is written as an
Ito stochastic term. In the numerical algorithm described in Section
\ref{sec:NumericalAlgorithm}, we use a second-order tau leaping method
\cite{WeakTrapezoidal}, which discretizes (\ref{Rs}) with a finite
time step size $\D t$.

\section{\label{app:EntropyArgument}Electroneutral Fluctuations of Composition
for a Binary Electrolyte}

For a binary electroneutral electrolyte, the covariance of the fluctuations
of the two charged species (\ref{eq:S_ideal_eln}) is the matrix
\[
\M S_{\text{ions}}^{(\text{eln})}=\frac{\rho}{1+b}\left[\begin{array}{cc}
m_{1}w_{1} & bm_{2}w_{1}\\
bm_{2}w_{1} & bm_{2}w_{2}
\end{array}\right],
\]
where $b=-m_{1}z_{1}/m_{2}z_{2}=V_{1}/V_{2}$ is the ratio of the
number of atoms of the two species in one neutral salt molecule. It
is important to observe that this is \emph{not} what would be predicted
from a naive ambipolar approximation where one considers the two ions
to be bound and diffusing with the ambipolar diffusion coefficient
(\ref{eq:D_amb}), notably, such an approximation would not give the
prefactor $\left(1+b\right)^{-1}$.

One can understand the prefactor $\left(1+b\right)^{-1}$ by computing
the entropy of mixing of the solution under the constraint of charge
neutrality. Consider a dilute ideal solution of $N_{0}$ molecules
of a solvent species and $N_{1}\ll N_{0}$ molecules of one ion and
$N_{2}\ll N_{0}$ molecules of another counter-ion. For an electroneutral
mixture we have the constraint $N_{2}=bN_{1}$. The mixture has a
free energy of mixing
\[
\left(k_{B}T\right)^{-1}\D G_{\text{mix}}\approx N_{1}\left(\ln\frac{N_{1}}{N_{0}}-1\right)+N_{2}\left(\ln\frac{N_{2}}{N_{0}}-1\right)=N_{1}\left(\ln\frac{N_{1}}{N_{0}}-1\right)+bN_{1}\left(\ln\frac{bN_{1}}{N_{0}}-1\right).
\]
The second derivative of the free energy of mixing, which determines
the width of the Gaussian approximation of the entropy and thus the
inverse of $S_{11}$, is
\[
\left(\frac{\partial^{2}\D G_{\text{mix}}}{\partial N_{1}^{2}}\right)=\left(1+b\right)\frac{k_{B}T}{N_{1}},
\]
which has the additional prefactor $\left(1+b\right)$ relative to
the standard result without the electroneutrality constraint.

%\bibliographystyle{unsrt}
%\bibliography{References}

\end{document}